\documentclass[preprint]{aastex63}

\usepackage{natbib}
\usepackage{graphicx}
\usepackage{times}
\usepackage{apjfonts}
\usepackage{amsmath}
\usepackage{xcolor}
\usepackage{listings}
\usepackage{verbatim}
\usepackage{enumitem}
\usepackage{natbib}
\setcitestyle{citesep={,}}
\usepackage{lineno}

\providecommand\tablesize{\scriptsize}

\usepackage{soul}
\usepackage{amsmath}
\usepackage{amssymb}
\usepackage{xspace}
\usepackage{xifthen}
\usepackage{eso-pic}

\mathchardef\mhyphen="2D

\newlength{\dhatheight}

\newcommand{\code}[1]{\texttt{#1}\xspace}

\newcommand{\unit}[1]{\ensuremath{\mathrm{\,#1}}\xspace}

\newcommand{\e}{\unit{e^{-}}}

\newcommand{\bandvar}[2][]{
  \ifthenelse{\isempty{#1}}{\var{#2}}{\var{#2\_#1}}
}

\newcommand{\HEALPix}{\code{HEALPix}}
\newcommand{\healpix}{\HEALPix}

\newcommand{\var}[1]{\ensuremath{\texttt{\MakeUppercase{#1}}}\xspace}

\providecommand\physrep{\ref@jnl{Phys.~Rep.}}
\providecommand\apjs{\ref@jnl{ApJS}}
\providecommand{\jcap}{\ref@jnl{JCAP}}

\setwatermarkfontsize{0.75in}

\shorttitle{DES Six Year Calibration Stars}
\shortauthors{Rykoff {\it et al.}}

\defcitealias{2018AJ....155...41B}{Burke, Rykoff, et al., 2018}

\begin{document}

\reportnum{DES-2023-0766}
\reportnum{FERMILAB-TM-2784-PPD-SCD}


\title{The Dark Energy Survey Six-Year Calibration Star Catalog}


\author[0000-0001-9376-3135]{E.~S.~Rykoff}
\affiliation{Kavli Institute for Particle Astrophysics \& Cosmology, P. O. Box 2450, Stanford University, Stanford, CA 94305, USA}
\affiliation{SLAC National Accelerator Laboratory, Menlo Park, CA 94025, USA}

\author[0000-0001-7211-5729]{D.~L.~Tucker}
\affiliation{Fermi National Accelerator Laboratory, P. O. Box 500, Batavia, IL 60510, USA}

\author{D.~L.~Burke}
\affiliation{Kavli Institute for Particle Astrophysics \& Cosmology, P. O. Box 2450, Stanford University, Stanford, CA 94305, USA}
\affiliation{SLAC National Accelerator Laboratory, Menlo Park, CA 94025, USA}

\author[0000-0002-7069-7857]{S.~S.~Allam}
\affiliation{Fermi National Accelerator Laboratory, P. O. Box 500, Batavia, IL 60510, USA}

\author[0000-0001-8156-0429]{K.~Bechtol}
\affiliation{Physics Department, 2320 Chamberlin Hall, University of Wisconsin-Madison, 1150 University Avenue Madison, WI  53706-1390}

\author[0000-0002-8613-8259]{G.~M.~Bernstein}
\affiliation{Department of Physics and Astronomy, University of Pennsylvania, Philadelphia, PA 19104, USA}

\author[0000-0001-5201-8374]{D.~Brout}
\affiliation{Center for Astrophysics $\vert$ Harvard \& Smithsonian, 60 Garden Street, Cambridge, MA 02138, USA}

\author[0000-0002-4588-6517]{R.~A.~Gruendl}
\affiliation{Center for Astrophysical Surveys, National Center for Supercomputing Applications, 1205 West Clark St., Urbana, IL 61801, USA}
\affiliation{Department of Astronomy, University of Illinois at Urbana-Champaign, 1002 W. Green Street, Urbana, IL 61801, USA}

\author[0000-0003-2999-4873]{J.~Lasker}
\affiliation{Department of Physics, Southern Methodist University, 3215 Daniel Avenue, Dallas, TX 75275, USA}
\affiliation{Kavli Institute for Cosmological Physics, University of Chicago, Chicago, IL 60637, USA}

\author[0000-0002-6261-4601]{J.~A.~Smith}
\affiliation{Dept. Physics, Engineering \& Astronomy, Austin Peay State University, Clarksville, TN  7044, USA}

\author[0000-0003-0072-6736]{W.~C.~Wester}
\affiliation{Fermi National Accelerator Laboratory, P. O. Box 500, Batavia, IL 60510, USA}

\author[0000-0002-9541-2678]{B.~Yanny}
\affiliation{Fermi National Accelerator Laboratory, P. O. Box 500, Batavia, IL 60510, USA}

\author[0000-0003-1587-3931]{T.~M.~C.~Abbott}
\affiliation{Cerro Tololo Inter-American Observatory, NSF's National Optical-Infrared Astronomy Research Laboratory, Casilla 603, La Serena, Chile}

\author[0000-0001-5679-6747]{M.~Aguena}
\affiliation{Laborat\'orio Interinstitucional de e-Astronomia - LIneA, Rua Gal. Jos\'e Cristino 77, Rio de Janeiro, RJ - 20921-400, Brazil}

\author[0000-0002-7394-9466]{O.~Alves}
\affiliation{Department of Physics, University of Michigan, Ann Arbor, MI 48109, USA}

\author[0000-0001-0171-6900]{F.~Andrade-Oliveira}
\affiliation{Department of Physics \& Astronomy, University College London, Gower Street, London, WC1E 6BT, UK}

\author[0000-0002-0609-3987]{J.~Annis}
\affiliation{Fermi National Accelerator Laboratory, P. O. Box 500, Batavia, IL 60510, USA}

\author[0000-0002-2562-8537]{D.~Bacon}
\affiliation{Institute of Cosmology and Gravitation, University of Portsmouth, Portsmouth, PO1 3FX, UK}

\author[0000-0002-3602-3664]{E.~Bertin}
\affiliation{CNRS, UMR 7095, Institut d'Astrophysique de Paris, F-75014, Paris, France}
\affiliation{Sorbonne Universit\'es, UPMC Univ Paris 06, UMR 7095, Institut d'Astrophysique de Paris, F-75014, Paris, France}

\author[0000-0002-8458-5047]{D.~Brooks}
\affiliation{Department of Physics \& Astronomy, University College London, Gower Street, London, WC1E 6BT, UK}

\author[0000-0003-3044-5150]{A.~Carnero~Rosell}
\affiliation{Instituto de Astrofisica de Canarias, E-38205 La Laguna, Tenerife, Spain}
\affiliation{Laborat\'orio Interinstitucional de e-Astronomia - LIneA, Rua Gal. Jos\'e Cristino 77, Rio de Janeiro, RJ - 20921-400, Brazil}
\affiliation{Universidad de La Laguna, Dpto. Astrofísica, E-38206 La Laguna, Tenerife, Spain}

\author[0000-0002-3130-0204]{J.~Carretero}
\affiliation{Institut de F\'isica d’Altes Energies (IFAE), The Barcelona Institute of Science and Technology, Campus UAB, 08193 Bellaterra (Barcelona), Spain}

\author[0000-0001-7316-4573]{F.~J.~Castander}
\affiliation{Institut d'Estudis Espacials de Catalunya (IEEC), 08034 Barcelona, Spain}
\affiliation{Institute of Space Sciences (ICE, CSIC),  Campus UAB, Carrer de Can Magrans, s/n,  08193 Barcelona, Spain}

\author[0000-0002-5636-233X]{A.~Choi}
\affiliation{NASA Goddard Space Flight Center, 8800 Greenbelt Rd, Greenbelt, MD 20771, USA}

\author[0000-0002-7731-277X]{L.~N.~da Costa}
\affiliation{Laborat\'orio Interinstitucional de e-Astronomia - LIneA, Rua Gal. Jos\'e Cristino 77, Rio de Janeiro, RJ - 20921-400, Brazil}

\author[0000-0002-7131-7684]{M.~E.~S.~Pereira}
\affiliation{Hamburger Sternwarte, Universit\"{a}t Hamburg, Gojenbergsweg 112, 21029 Hamburg, Germany}

\author[0000-0002-4213-8783]{T.~M.~Davis}
\affiliation{School of Mathematics and Physics, University of Queensland,  Brisbane, QLD 4072, Australia}

\author[0000-0001-8318-6813]{J.~De~Vicente}
\affiliation{Centro de Investigaciones Energ\'eticas, Medioambientales y Tecnol\'ogicas (CIEMAT), Madrid, Spain}

\author[0000-0002-8357-7467]{H.~T.~Diehl}
\affiliation{Fermi National Accelerator Laboratory, P. O. Box 500, Batavia, IL 60510, USA}

\author{P.~Doel}
\affiliation{Department of Physics \& Astronomy, University College London, Gower Street, London, WC1E 6BT, UK}

\author[0000-0001-8251-933X]{A.~Drlica-Wagner}
\affiliation{Department of Astronomy and Astrophysics, University of Chicago, Chicago, IL 60637, USA}
\affiliation{Fermi National Accelerator Laboratory, P. O. Box 500, Batavia, IL 60510, USA}
\affiliation{Kavli Institute for Cosmological Physics, University of Chicago, Chicago, IL 60637, USA}

\author[0000-0002-3745-2882]{S.~Everett}
\affiliation{Jet Propulsion Laboratory, California Institute of Technology, 4800 Oak Grove Dr., Pasadena, CA 91109, USA}

\author{I.~Ferrero}
\affiliation{Institute of Theoretical Astrophysics, University of Oslo. P.O. Box 1029 Blindern, NO-0315 Oslo, Norway}

\author[0000-0003-4079-3263]{J.~Frieman}
\affiliation{Fermi National Accelerator Laboratory, P. O. Box 500, Batavia, IL 60510, USA}
\affiliation{Department of Astronomy and Astrophysics, University of Chicago, Chicago, IL 60637, USA}
\affiliation{Kavli Institute for Cosmological Physics, University of Chicago, Chicago, IL 60637, USA}

\author[0000-0002-9370-8360]{J.~Garc\'ia-Bellido}
\affiliation{Instituto de Fisica Teorica UAM/CSIC, Universidad Autonoma de Madrid, 28049 Madrid, Spain}

\author[0000-0002-3730-1750]{G.~Giannini}
\affiliation{Institut de F\'isica d’Altes Energies (IFAE), The Barcelona Institute of Science and Technology, Campus UAB, 08193 Bellaterra (Barcelona), Spain}

\author[0000-0003-3270-7644]{D.~Gruen}
\affiliation{University Observatory, Faculty of Physics, Ludwig-Maximilians-Universit\"at, Scheinerstr. 1, 81679 Munich, Germany}

\author[0000-0003-0825-0517]{G.~Gutierrez}
\affiliation{Fermi National Accelerator Laboratory, P. O. Box 500, Batavia, IL 60510, USA}

\author[0000-0003-2071-9349]{S.~R.~Hinton}
\affiliation{School of Mathematics and Physics, University of Queensland,  Brisbane, QLD 4072, Australia}

\author[0000-0002-9369-4157]{D.~L.~Hollowood}
\affiliation{Santa Cruz Institute for Particle Physics, Santa Cruz, CA 95064, USA}

\author[0000-0001-5160-4486]{D.~J.~James}
\affiliation{Center for Astrophysics $\vert$ Harvard \& Smithsonian, 60 Garden Street, Cambridge, MA 02138, USA}

\author[0000-0003-0120-0808]{K.~Kuehn}
\affiliation{Australian Astronomical Optics, Macquarie University, North Ryde, NSW 2113, Australia}
\affiliation{Lowell Observatory, 1400 Mars Hill Rd, Flagstaff, AZ 86001, USA}

\author[0000-0002-1134-9035]{O.~Lahav}
\affiliation{Department of Physics \& Astronomy, University College London, Gower Street, London, WC1E 6BT, UK}

\author[0000-0003-0710-9474]{J.~L.~Marshall}
\affiliation{George P. and Cynthia Woods Mitchell Institute for Fundamental Physics and Astronomy, and Department of Physics and Astronomy, Texas A\&M University, College Station, TX 77843,  USA}

\author[0000-0001-9497-7266]{J. Mena-Fern{\'a}ndez}
\affiliation{Centro de Investigaciones Energ\'eticas, Medioambientales y Tecnol\'ogicas (CIEMAT), Madrid, Spain}

\author[0000-0002-1372-2534]{F.~Menanteau}
\affiliation{Center for Astrophysical Surveys, National Center for Supercomputing Applications, 1205 West Clark St., Urbana, IL 61801, USA}
\affiliation{Department of Astronomy, University of Illinois at Urbana-Champaign, 1002 W. Green Street, Urbana, IL 61801, USA}

\author[0000-0001-6145-5859]{J.~Myles}
\affiliation{Department of Physics, Stanford University, 382 Via Pueblo Mall, Stanford, CA 94305, USA}
\affiliation{Kavli Institute for Particle Astrophysics \& Cosmology, P. O. Box 2450, Stanford University, Stanford, CA 94305, USA}
\affiliation{SLAC National Accelerator Laboratory, Menlo Park, CA 94025, USA}

\author[0000-0001-6706-8972]{B.~D.~Nord}
\affiliation{Fermi National Accelerator Laboratory, P. O. Box 500, Batavia, IL 60510, USA}
\affiliation{Department of Astronomy and Astrophysics, University of Chicago, Chicago, IL 60637, USA}
\affiliation{Kavli Institute for Cosmological Physics, University of Chicago, Chicago, IL 60637, USA}
\affiliation{Laboratory for Nuclear Science, MIT, Cambridge MA, 02139, USA}

\author[0000-0003-2120-1154]{R.~L.~C.~Ogando}
\affiliation{Observat\'orio Nacional, Rua Gal. Jos\'e Cristino 77, Rio de Janeiro, RJ - 20921-400, Brazil}

\author[0000-0002-6011-0530]{A.~Palmese}
\affiliation{Department of Physics, Carnegie Mellon University, Pittsburgh, Pennsylvania 15312, USA}

\author[0000-0001-9186-6042]{A.~Pieres}
\affiliation{Laborat\'orio Interinstitucional de e-Astronomia - LIneA, Rua Gal. Jos\'e Cristino 77, Rio de Janeiro, RJ - 20921-400, Brazil}
\affiliation{Observat\'orio Nacional, Rua Gal. Jos\'e Cristino 77, Rio de Janeiro, RJ - 20921-400, Brazil}

\author[0000-0002-2598-0514]{A.~A.~Plazas~Malag\'on}
\affiliation{Department of Astrophysical Sciences, Princeton University, Peyton Hall, Princeton, NJ 08544, USA}

\author{M.~Raveri}
\affiliation{Department of Physics, University of Genova and INFN, Via Dodecaneso 33, 16146, Genova, Italy}

\author[0000-0001-6163-1058]{M.~Rodgr\'iguez-Monroy}
\affiliation{Universit\'e Paris-Saclay, CNRS/IN2P3, IJCLab, 91405, Orsay, France}

\author[0000-0002-9646-8198]{E.~Sanchez}
\affiliation{Centro de Investigaciones Energ\'eticas, Medioambientales y Tecnol\'ogicas (CIEMAT), Madrid, Spain}

\author{B.~Santiago}
\affiliation{Instituto de F\'\i sica, UFRGS, Caixa Postal 15051, Porto Alegre, RS - 91501-970, Brazil}
\affiliation{Laborat\'orio Interinstitucional de e-Astronomia - LIneA, Rua Gal. Jos\'e Cristino 77, Rio de Janeiro, RJ - 20921-400, Brazil}

\author[0000-0001-9504-2059]{M.~Schubnell}
\affiliation{Department of Physics, University of Michigan, Ann Arbor, MI 48109, USA}

\author[0000-0002-1831-1953]{I.~Sevilla-Noarbe}
\affiliation{Centro de Investigaciones Energ\'eticas, Medioambientales y Tecnol\'ogicas (CIEMAT), Madrid, Spain}

\author[0000-0002-3321-1432]{M.~Smith}
\affiliation{School of Physics and Astronomy, University of Southampton,  Southampton, SO17 1BJ, UK}

\author[0000-0001-6082-8529]{M.~Soares-Santos}
\affiliation{Department of Physics, University of Michigan, Ann Arbor, MI 48109, USA}

\author[0000-0002-7047-9358]{E.~Suchyta}
\affiliation{Computer Science and Mathematics Division, Oak Ridge National Laboratory, Oak Ridge, TN 37831}

\author{M.~E.~C.~Swanson}
\affiliation{Center for Astrophysical Surveys, National Center for Supercomputing Applications, 1205 West Clark St., Urbana, IL 61801, USA}

\author{T.~N.~Varga}
\affiliation{Max Planck Institute for Extraterrestrial Physics, Giessenbachstrasse, 85748 Garching, Germany}
\affiliation{University Observatory, Faculty of Physics, Ludwig-Maximilians-Universit\"at, Scheinerstr. 1, 81679 Munich, Germany, Germany}

\author[0000-0001-8788-1688]{M.~Vincenzi}
\affiliation{Department of Physics, Duke University Durham, NC 27708, USA}

\author[0000-0002-7123-8943]{A.~R.~Walker}
\affiliation{Cerro Tololo Inter-American Observatory, NSF's National Optical-Infrared Astronomy Research Laboratory, Casilla 603, La Serena, Chile}

\author[0000-0001-9382-5199]{N.~Weaverdyck}
\affiliation{Department of Physics, University of Michigan, Ann Arbor, MI 48109, USA}
\affiliation{Lawrence Berkeley National Laboratory, 1 Cyclotron Road, Berkeley, CA 94720, USA}

\author[0000-0002-3073-1512]{P.~Wiseman}
\affiliation{School of Physics and Astronomy, University of Southampton,  Southampton, SO17 1BJ, UK}

\collaboration{121}{(DES Collaboration)}

\correspondingauthor{Eli Rykoff}
\email{erykoff@slac.stanford.edu}

\begin{abstract}


This Technical Note presents a catalog of calibrated reference stars that was generated by the Forward Calibration Method (FGCM) pipeline \citepalias{2018AJ....155...41B} as part of the FGCM photometric calibration of the full Dark Energy Survey (DES) 6-Year data set (Y6).  This catalog provides DES $grizY$ magnitudes for 17 million stars with $i$-band magnitudes mostly in the range $16 \la i \la 21$ spread over the full DES footprint covering $5000\,\mathrm{deg}^2$ over the Southern Galactic Cap at galactic latitudes $b \la -20^\circ$ (plus a few outlying fields disconnected from the main survey footprint). These stars are calibrated to a uniformity of better than $1.8~\mathrm{mmag}$ (0.18\%) RMS over the survey area.  The absolute calibration of the catalog is computed with reference to the \var{stisnic.007} spectrum of the Hubble Space Telescope CalSpec standard star C26202; including systematic errors, the absolute flux system is known at the $\approx 1\%$ level. As such, these stars provide a useful reference catalog for calibrating $grizY$-band or $grizY$-like band photometry in the Southern Hemisphere, particularly for observations within the DES footprint.

\end{abstract}


\section{Introduction}
\label{sec:intro}
\subsection{The Dark Energy Survey}
\label{subsec:DES}


The Dark Energy Survey (DES) began observations in 2013 using the Dark Energy Camera (DECam) mounted on the Victor M.\ Blanco 4-meter telescope at the Cerro Tololo Inter-American Observatory (CTIO) \citep{2010JPhCS.259a2080S, Diehl:2014, Diehl:2019}.  The DECam instrument \citep{2015AJ....150..150F} is a 570 megapixel CCD research camera composed of 74 individual CCDs, with active field of view of 2.7 deg$^{2}$ at the f/2.7 prime focus of the Blanco.  Science data were collected using the DES $grizY$ filters.  When completed in 2019 the survey covered approximately $5000\,\mathrm{deg}^{2}$ of the southern hemisphere sky with 8-10 repeated observations in each band taken over a period of 6 years~\citep{2019arXiv191206254N}, comprising the full 6-year data set (Y6).  The data were processed by the DES data management system \citep{2012ApJ...757...83D} at the National Center for Supercomputing Applications at the University of Illinois.  A complete description of the DES image processing pipeline can be found in \citet{DESDM}.  This pipeline utilizes image quality cuts and processing based on instrumental throughput derived from dedicated dithered observations of selected star fields~\citep[see][]{2018PASP..130e4501B}.

\subsection{Overview Of The DES Photometric Calibration}
\label{subsec:DES}


Photometric calibration of the DES broadband images is performed using the Forward Global Calibration Method (FGCM), which is described in detail in \citetalias{2018AJ....155...41B}\footnote{https://github.com/erykoff/fgcm/tree/v3.3.1/fgcm} and further updated in \cite{DESDR2}.  FGCM calibrates the full survey with a forward model that incorporates the known system throughput (including positional variation in the transmission curve of filters and chromatic quantum efficiency differences in the detectors) with a model of the atmosphere.  Isolated stars (with no neighbor within $2"$) are selected from single-epoch images, and all calibration stars are observed with a signal-to-noise greater than 10 in at least 2 observations in each of $griz$.  The PSF flux is used as the calibration flux to minimize issues in photometry for faint stars due to offsets in the local background.  Stars with broad range of colors ($0.5 < g - i < 3.5$) are used to constrain atmospheric parameters in a full multi-band calibration.  A final down-sampling of stars in high density regions is performed prior to performing the calibration fit to ensure that the overall chi-squared is not dominated by the edges of the survey.

The FGCM fit is iterative, with the model based on "photometric" observations that are defined as those that have fluxes consistent with variations in the atmosphere that can be described by variations in aerosol, precipitable water vapor, and ozone, in addition to the variations in airmass predicted from the MODTRAN model~\citep{1999SPIE.3756..348B}, which is described fully in \citetalias{2018AJ....155...41B}.  We note that 80\% of the full set of exposures are considered photometric, in that variations can be described by the atmosphere model without excess gray extinction or a significant increase in variance across the focal plane.  A number of improvements to the model have been made since the original FGCM paper, including (a) treatment of aperture effects caused by variations in the wings of the PSF~\citep[see, e.g.,][]{2018PASP..130e4501B}; (b) a linear approximation to the chromatic degradation of the primary mirror surface between recoatings of the mirror; (c) the GPS data is no longer used as a constraint on the atmospheric water vapor.

After the convergence of the model fit, non-photometric exposures are matched to the preliminary network of stars to produce zero-points and chromatic correction terms for every CCD/exposure in the survey, provided they have sufficient overlap with the calibration stars.  We then produce the final DES Year 6 Calibration Star Catalog presented here by applying these calibration parameters to every star observation, including ones that were not used in the fit due to being in higher density regions.

\section{Catalog Description}
\label{sec:cat_descript}

\begin{figure*}[h]
\centering
\includegraphics[scale=1.0]{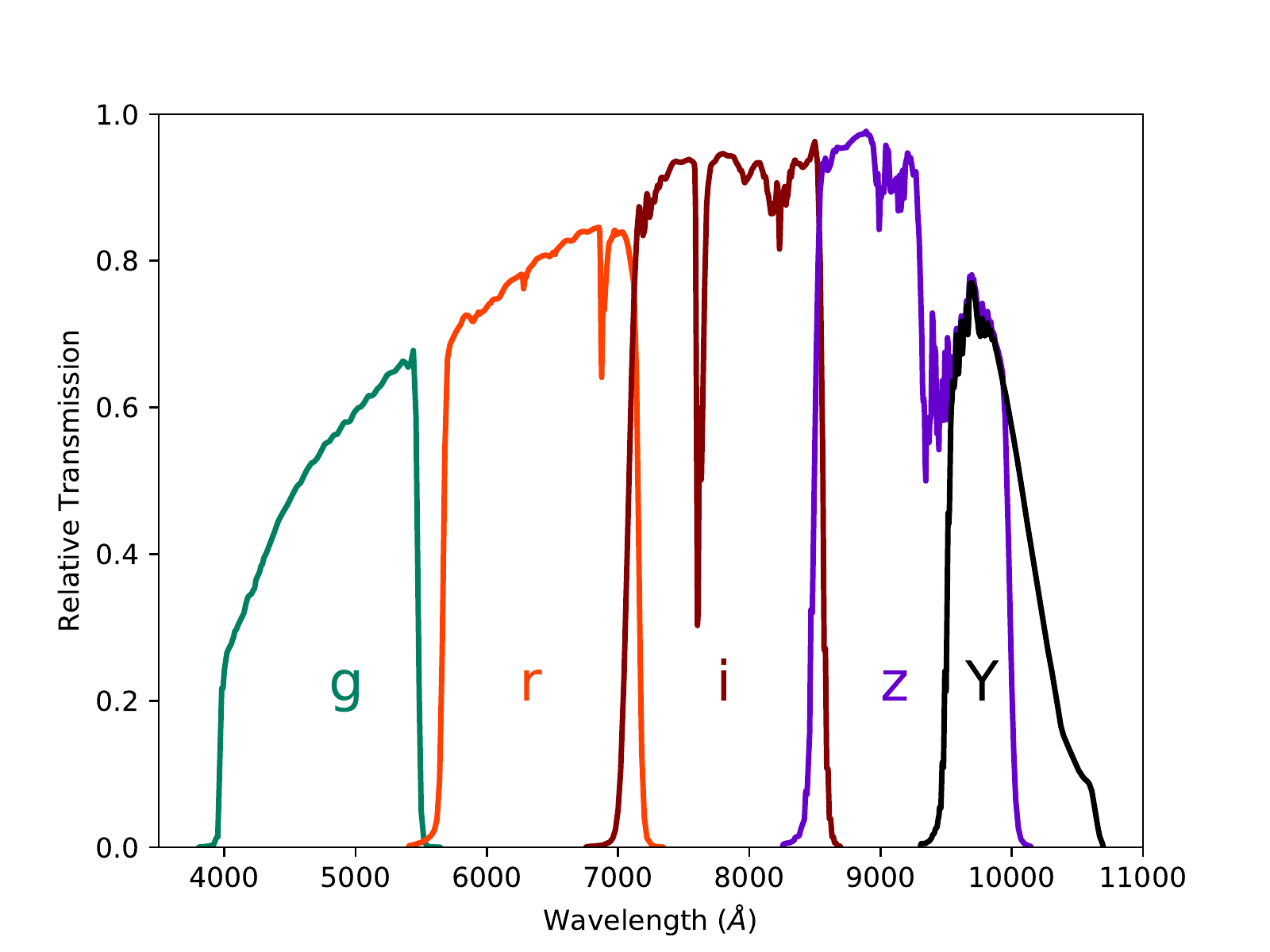}
\caption{A plot of the DES standard bandpass curves ($grizY$) used in DES DR2 \citep{DESDR2}.
The bandpasses represent the total system throughput, including atmospheric transmission (air mass = 1.2) and the average instrumental response across the science CCDs.  Machine-readable tables of the DES standard bandpasses can be found at \url{https://des.ncsa.illinois.edu/releases/dr2/dr2-products}}
\label{fig:DESDR2_std_bandpasses}
\end{figure*}

In this Technical Note, we wish merely to provide this DES 6-Year (Y6) Calibration Star Catalog and to describe its general characteristics in order to facilitate its use as a photometric reference catalog for the calibration of other data sets using similar filter bandpasses.  As noted in the previous section, in creating this reference catalog, FGCM took into account and and removed the effects of positional (across the DECam CCD mosaic) and temporal variations of the effective bandpasses for a observations on a given CCD at at given time and date; the photometry has been calibrated to the DES standard bandpasses, which are shown in Figure~\ref{fig:DESDR2_std_bandpasses}.  The full catalog contains roughly 17 million stars with $i$-band magnitudes mostly in the range $16 \la i \la 21$.

\begin{figure*}[h]
\centering
\includegraphics[scale=0.60]{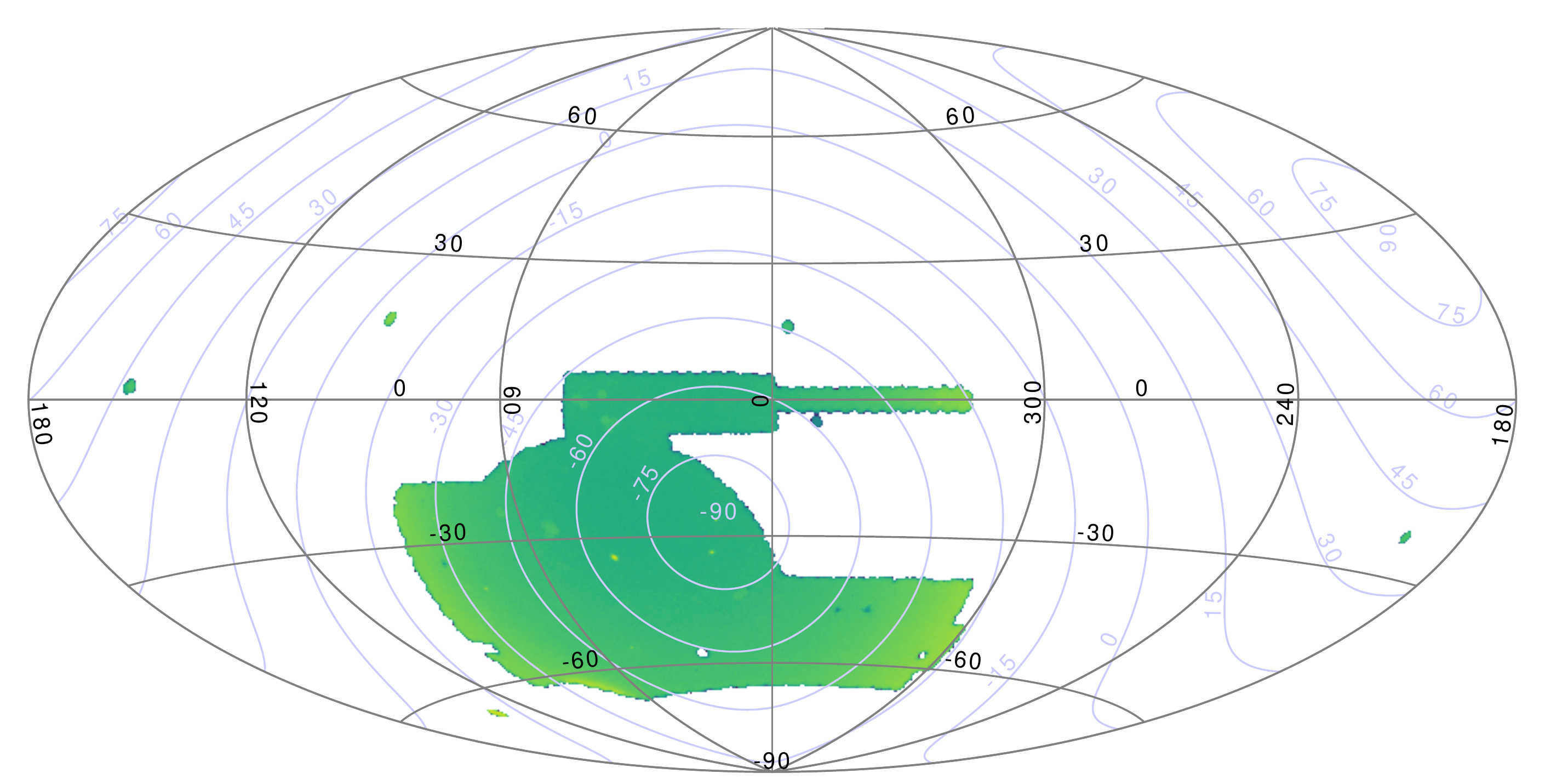}
\caption{A (logarithmically scaled) sky density Aitoff projection of the full DES Y6 FGCM calibration star catalog, plotted in equatorial coordinates (black grid).  The light purple grid lines indicate contours of galactic latitude, $b$. Note that the main contiguous footprint lies within $b\la-20~deg$, although there are a handful of fields outside this main footprint.}
\label{fig:aitoff_density}
\end{figure*}

In Figure~\ref{fig:aitoff_density}, we show the sky density distribution of the stars in the this catalog.  Nearly all fall within the official contiguous DES footprint, but we note that there are also some outlier fields that may be of interest to the user of this catalog.

\begin{figure*}[h]
\centering
\includegraphics[scale=0.50]{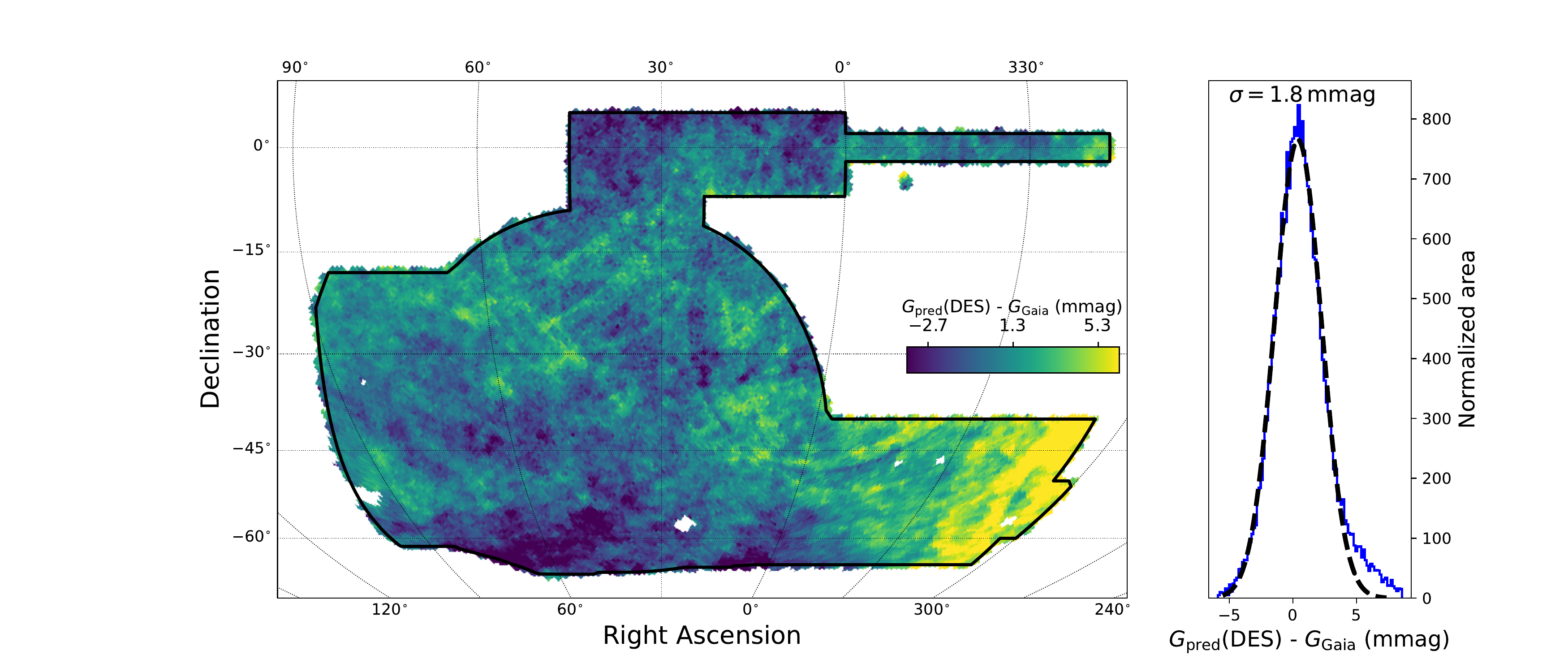}
\caption{\emph{Left:} A map of the per-pixel average offset between Gaia G-band fluxes synthesized from DES $griz$ fluxes and the measured Gaia G-band fluxes (HEALPix NSIDE=128).  We note that there is an offset of up to 1\% at the edge of the survey toward the Galactic Bulge (comprising about 5\% of the area).  We have yet to determine if this is an issue with the DES calibrations, the Gaia calibrations, Galactic reddening residuals, background issues, or some combination of these. \emph{Right:} Histogram of the per-pixel offset between these maps.}
\label{fig:gaia_comparison}
\end{figure*}

In Figure~\ref{fig:gaia_comparison}, we plot the offset between the Gaia $G$-band synthesized magnitudes transformed from the DES $griz$ magnitudes ($G_{\rm pred}$) of the stars in this catalog and their true measured Gaia $G$-band magnitude ($G_{\rm Gaia}$) from the Gaia Data Release 3 (Gaia DR3; \citep{GaiaDR3_Summary_2022}).  The synthesized $G$-band magnitudes were calculated in a manner similar to that in \cite{DESDR2}.  For this comparison, stars were selected to have $0.5 < g-i < 1.5$ and $G < 20$. Note that the RMS in in the difference in the synthesized and genuine $G$-band magnitudes ($G_{\rm pred}-G_{\rm Gaia}$), averaged over an NSIDE$=$128 HEALPix pixel (sky area $\approx$ 0.21~sq~deg), is a mere 1.8~mmag (0.018\%), and the main coherent features in the map appear to be of Gaia scan patterns, not DES hexagonal pointings.  Based on this plot, we estimate that the FGCM calibration star photometry is uniform across the DES footprint at the 1.8~mmag (0.18\%) RMS level or better.  We note that there is an offset of up to 1\% at the edge of the survey toward the Galactic Bulge (comprising about 5\% of the area).  We have yet to determine if this is an issue with the DES calibrations, the Gaia calibrations, Galactic reddening residuals, background issues, or some combination of these.  Overall, we attribute the exceptional uniformity of the calibration to the stability of DECam as well as the simplicity of the survey strategy, with all wide-field (WF) survey exposures of an equal 90 seconds.

\begin{figure*}[h]
\centering
\includegraphics[scale=0.70]{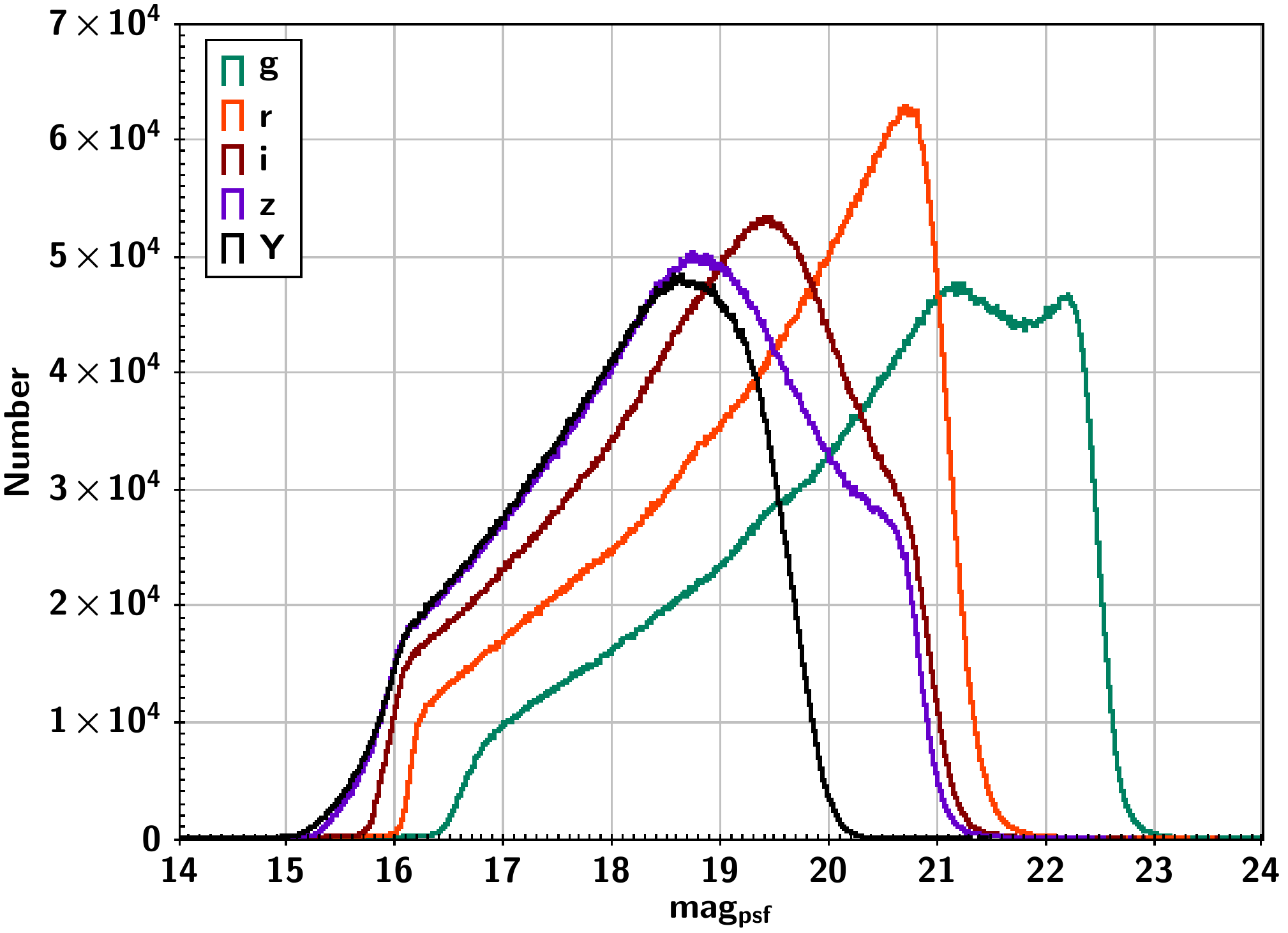}
\caption{Histogram of the distribution of magnitudes for the standard star catalog in each of the 5 DES filter bands.}
\label{fig:mag_hist}
\end{figure*}

For a calibration reference catalog to be useful, one needs to know over which magnitude ranges it covers.  As we see in Figure~\ref{fig:mag_hist}, this catalog roughly covers a range in $i$-band of 16--21.  Although there are a small fraction of stars with $i$-band outside this range, the vast majority lie within this range.  Other bands have somewhat different ranges, due both to the typical colors of stars and to the overall sensitivity of the DES survey in these other bands.

\begin{figure*}[h]
\includegraphics[scale=0.70]{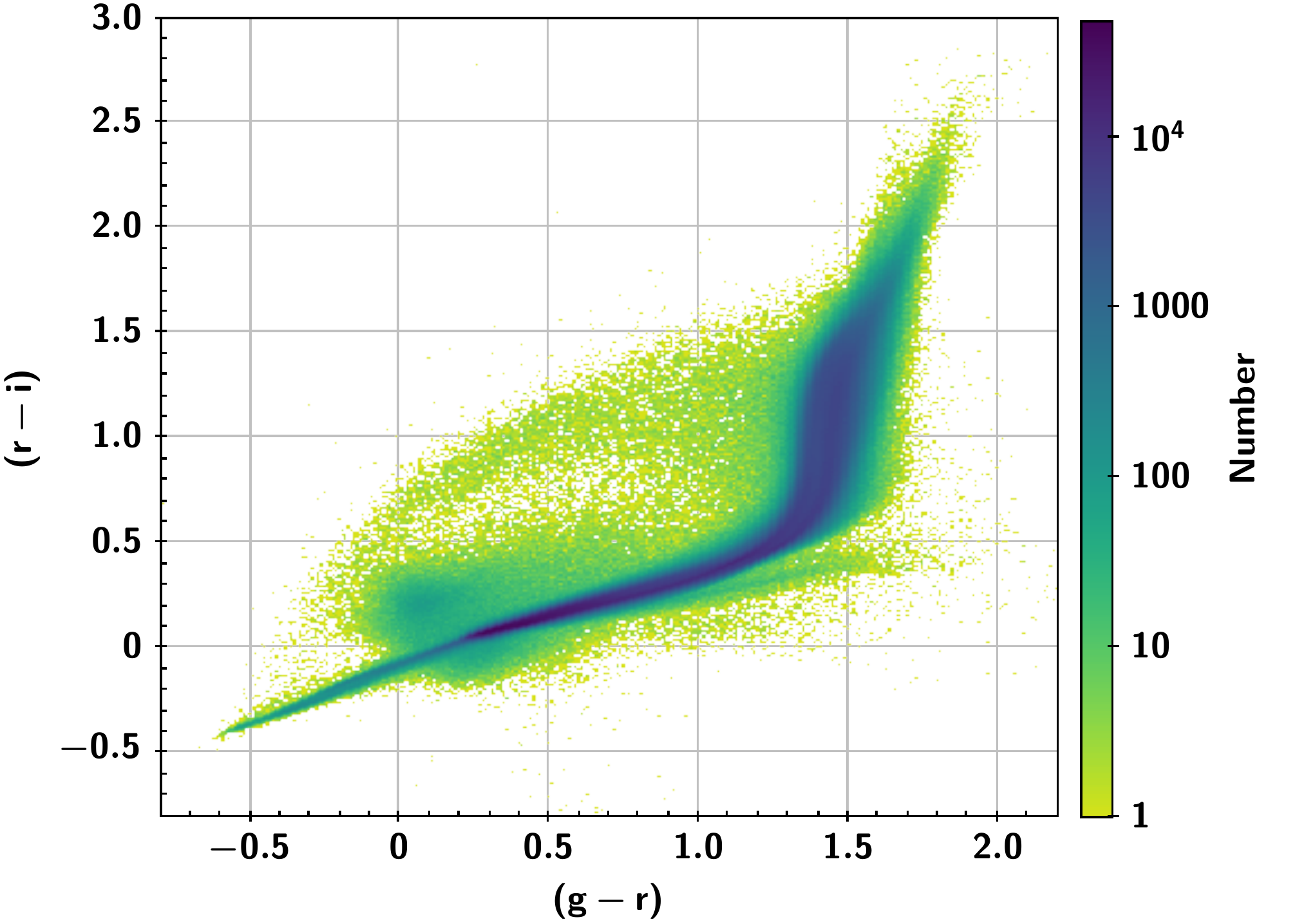}
\includegraphics[scale=0.70]{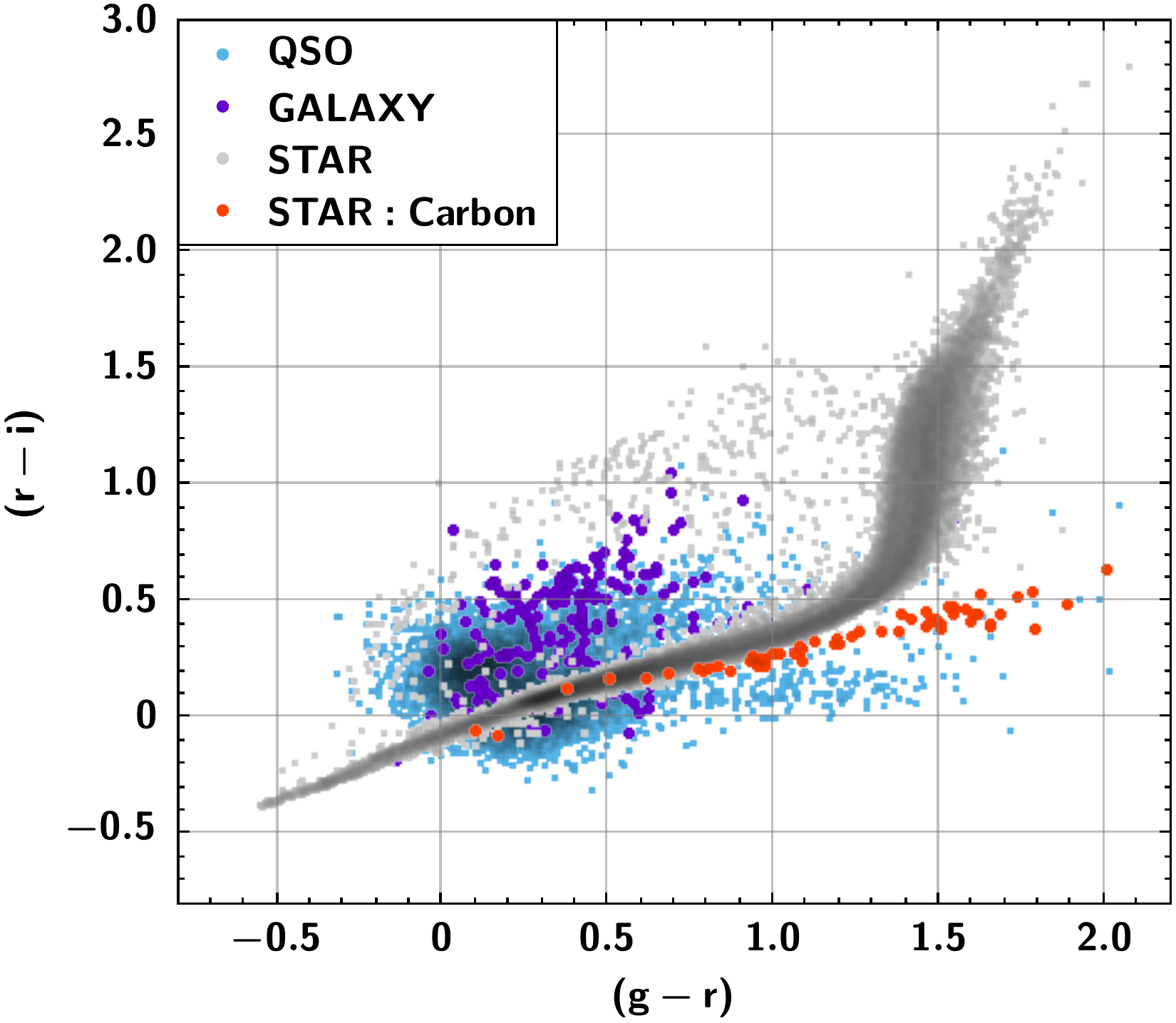}
\caption{{\em (top)} The $g-r$, $r-i$ color-color diagram for all stars with $N_{\rm good} > 2$ for the $g$, $r$, $i$ bandpasses.\\
{\em (bottom)} Same as the top figure, but just for those FGCM calibration stars with matches to the \texttt{SpecPhoto} view from the Sloan Digital Sky Survey Data Release 17 (SDSS DR17; \citealt{SDSS_DR17}) database available on the SDSS CasJobs server (\url{https://skyserver.sdss.org/CasJobs}).  In the \texttt{SpecPhoto} view, objects depicted as \textcolor{gray}{gray symbols} were classified as stars; \textcolor{cyan}{cyan symbols} are objects that were classified as Quasi-Stellar Objects (QSOs); \textcolor{purple}{purple symbols} were classified as galaxies; and \textcolor{red}{red symbols} were stars that were further subclassified as carbon stars.}
\label{fig:gri}
\end{figure*}

\begin{figure*}[h]
\includegraphics[scale=0.70]{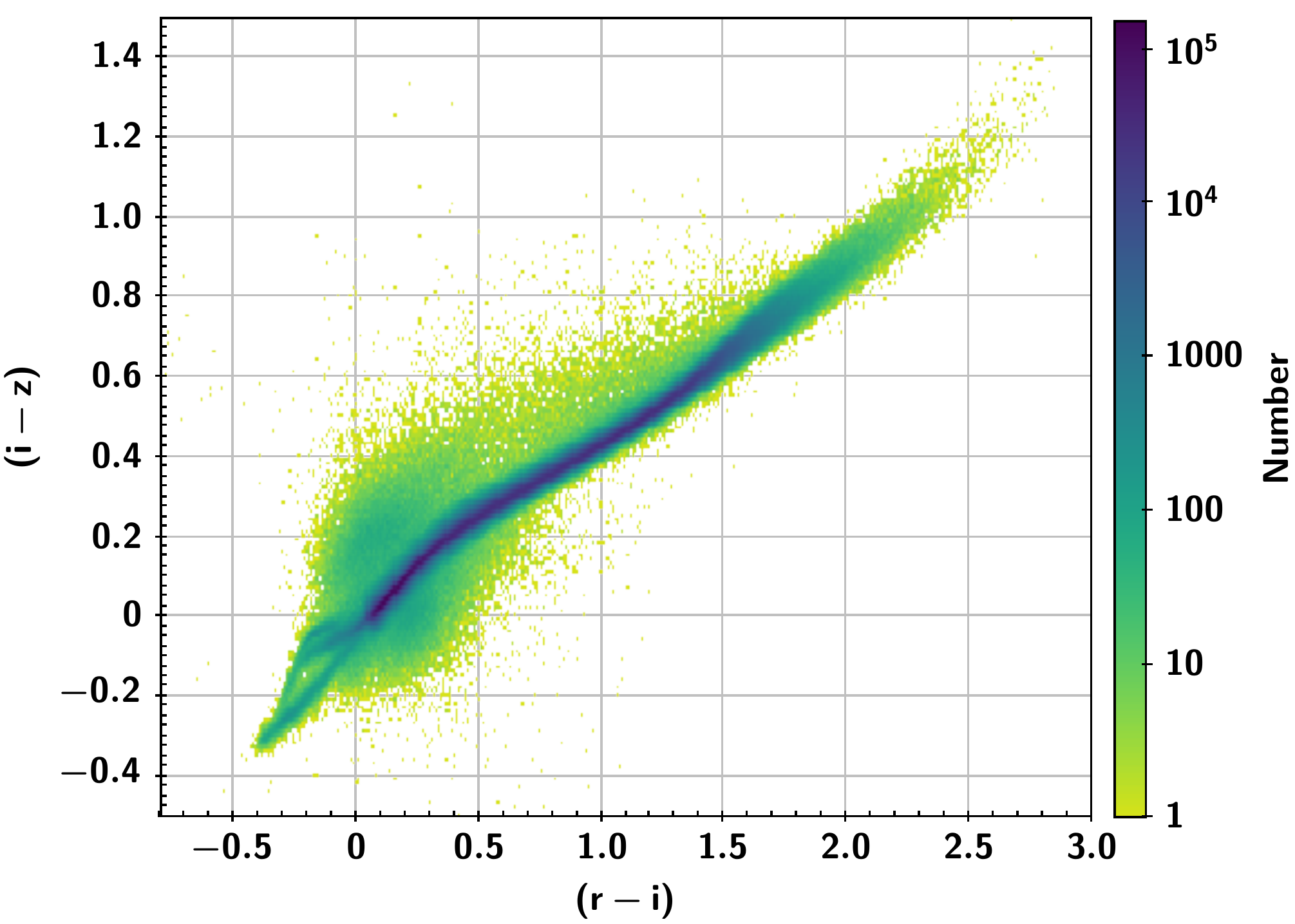}
\vspace{1.0cm}
\includegraphics[scale=0.70]{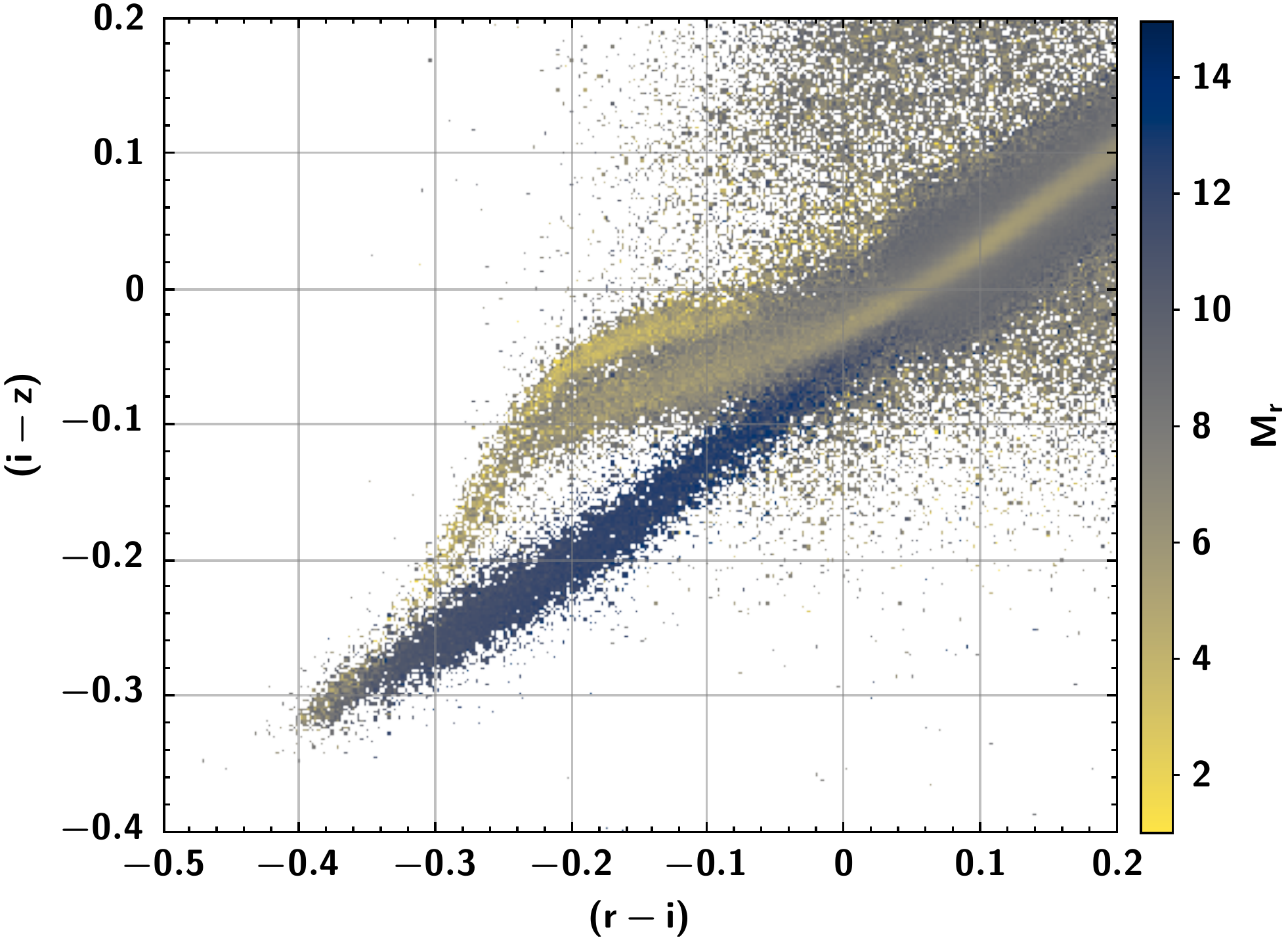}
\caption{{\em (top)} The $r-i$, $i-z$ color-color diagram for all stars with $N_{\rm good} > 2$ for the $r$, $i$, $z$ bandpasses.\\
{\em (bottom)} Same as top figure, but zoomed onto the blue-blue quadrant and now color-coded by the $r$-band absolute magnitude ($M_r$) of each object, making use of the Gaia EDR3 geometric distances from \cite{2021AJ....161..147B}; roughly 90\% of FGCM calibration stars have Gaia EDR3 geometric distances.  Note the trifurcation of the stellar locus is composed of low-luminosity ($M_r\approx14$) white dwarf stars, medium-luminosity ($M_r\approx6$) main-sequence stars, and high-luminosity ($M_r\approx3$) blue horizontal branch (BHB) stars.  We note that \citet{2012AJ....143...86V} have previously used SDSS $z$-band as a proxy measure for stellar surface gravity to discrimate BHB stars from white dwarfs.}
\label{fig:riz}
\end{figure*}

\begin{figure*}[h]
\centering
\includegraphics[scale=0.70]{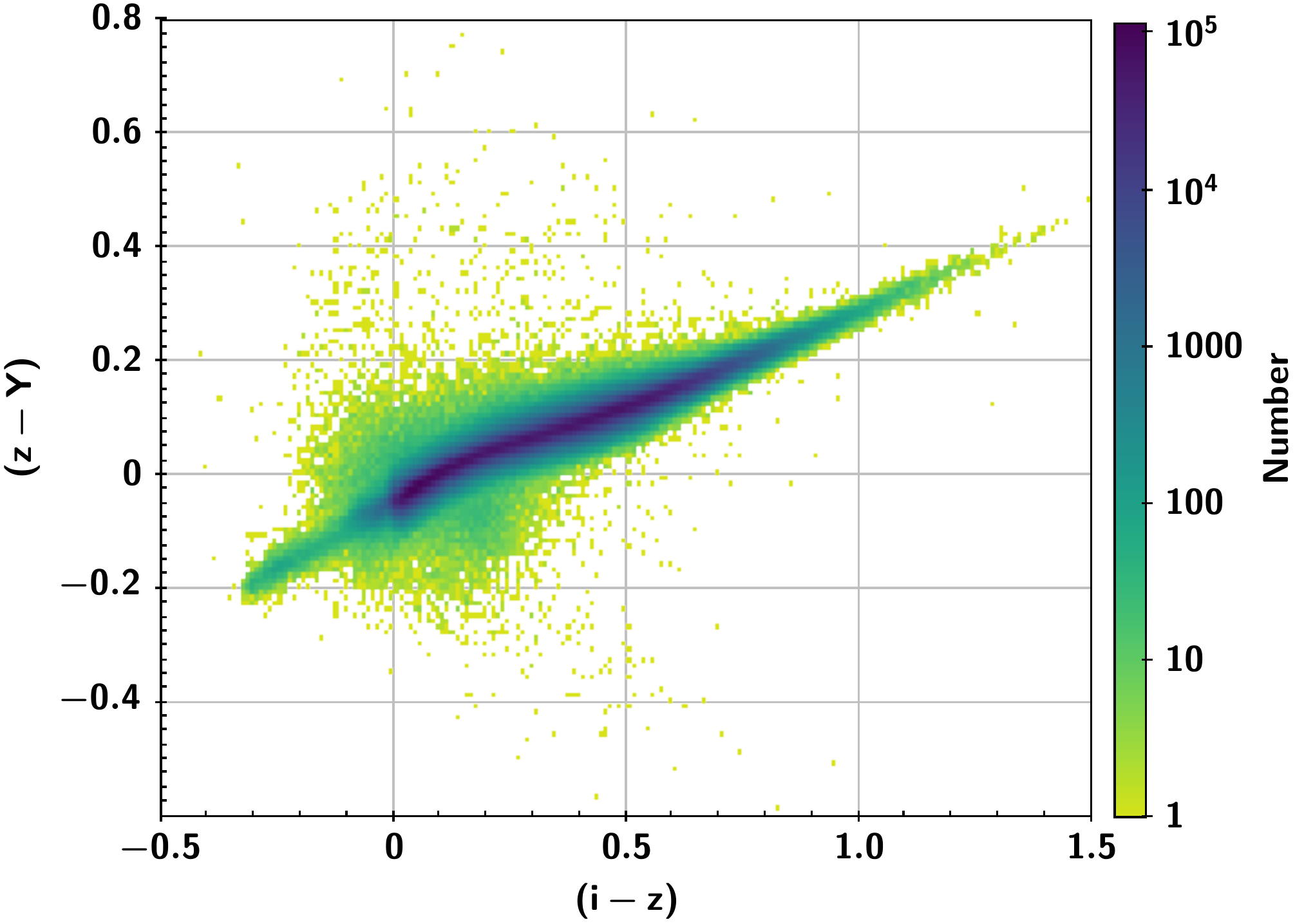}
\caption{The $i-z$, $z-Y$ color-color diagram for all stars with $N_{\rm good} > 2$ for the $i$, $z$, $Y$ bandpasses.}
\label{fig:izY}
\end{figure*}

In Figures~\ref{fig:gri}, \ref{fig:riz}, and \ref{fig:izY}, we plot the color-color diagrams for this catalog.  (In Figures~\ref{fig:gri} and \ref{fig:riz}, we also note some special features in these particular color-color diagrams.)  Note, in each figure, we include only those stars that have more than 2 good DES observations in each of the 3 filter bandpasses represented (e.g., $N_{\rm good}>2$ for each of the $g$, $r$, and $i$ bandpasses in Fig.~\ref{fig:gri}).  {\em For the highest quality work, we recommend the user to implement these quality cuts when using this catalog.}  Note that catalog contains a full breadth of stellar types, from hot white dwarfs ($g-r \la 0$, $r-i \la 0$) to M stars ($g-r \approx 1.6$).  Also note the tightness of the features in the stellar locus in each of these plots, which indicative of the photometric quality of this catalog.

\begin{figure*}[h]
\centering
\includegraphics[scale=0.60]{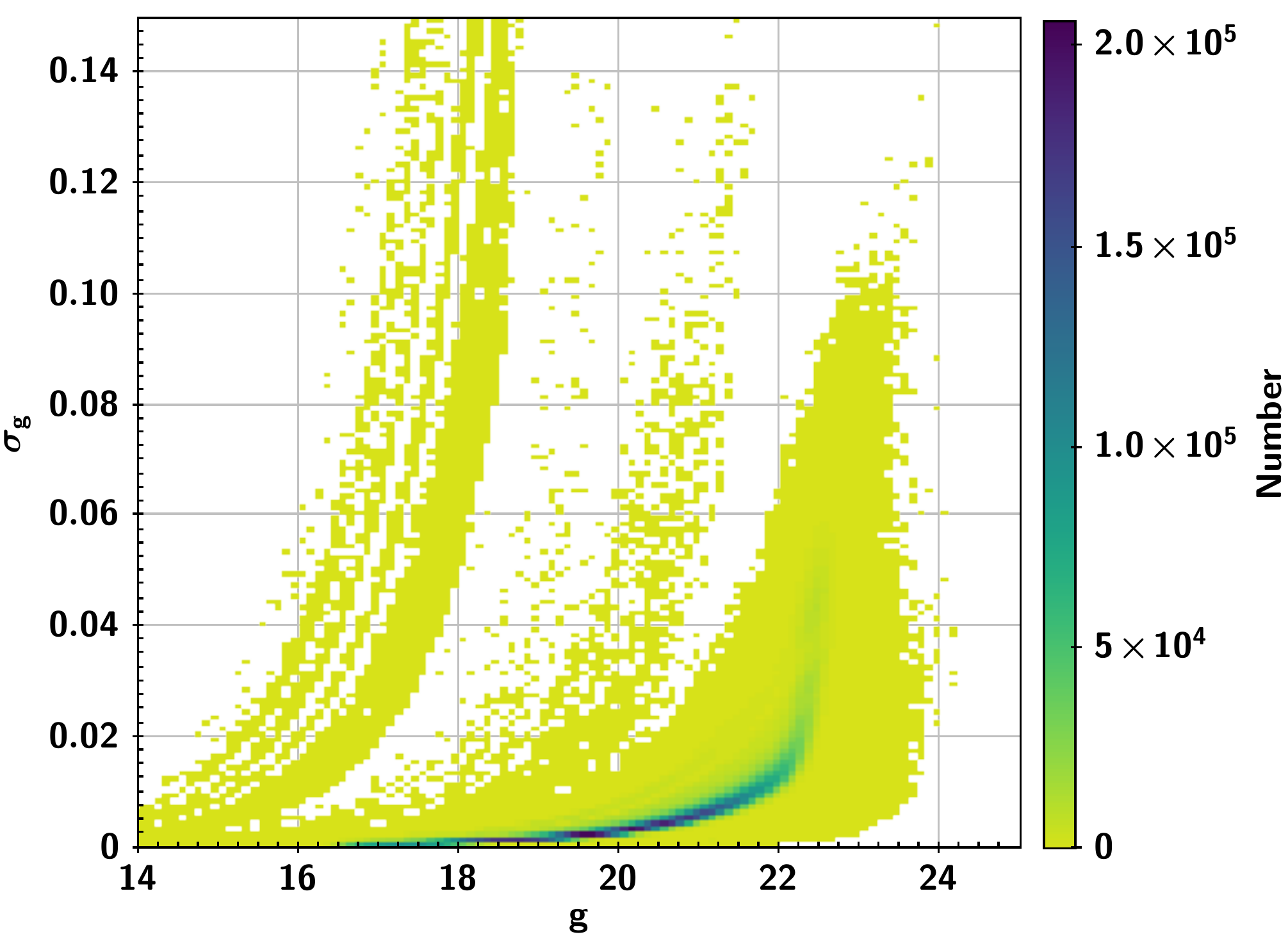}
\caption{$\sigma_g$ vs. $g$.  Data are binned in 0.001~mag ($\sigma_g$) $\times$ 0.1~mag ($g$) bins.  (Here we do not apply any $N_{\rm good}$ cuts.)}
\label{fig:magerr_vs_mag_g}
\end{figure*}

\begin{figure*}[h]
\centering
\includegraphics[scale=0.60]{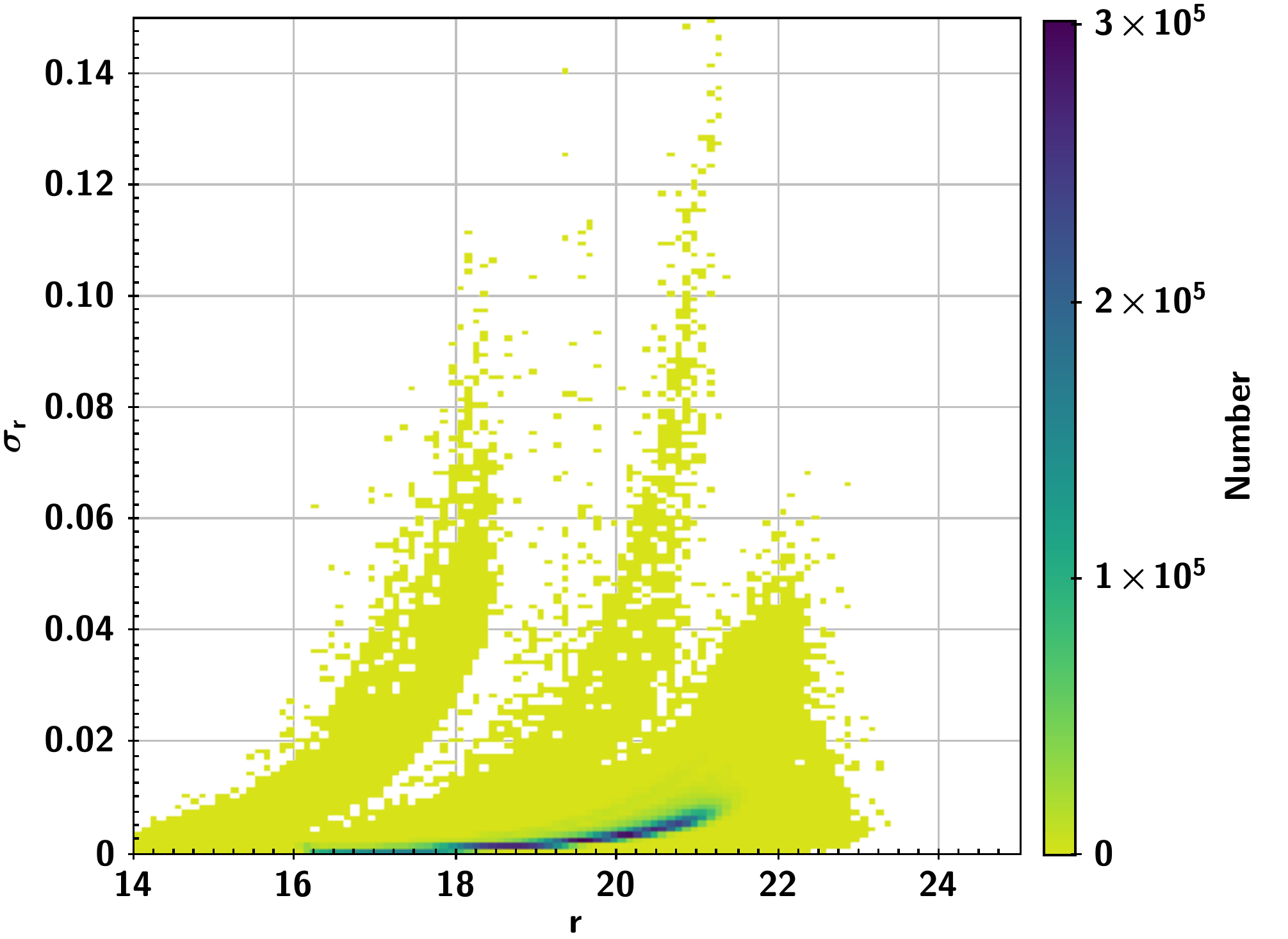}
\caption{$\sigma_r$ vs. $r$.  Data are binned in 0.001~mag ($\sigma_r$) $\times$ 0.1~mag ($r$) bins.  (Here we do not apply any $N_{\rm good}$ cuts.)}
\label{fig:magerr_vs_mag_r}
\end{figure*}

\begin{figure*}[h]
\centering
\includegraphics[scale=0.60]{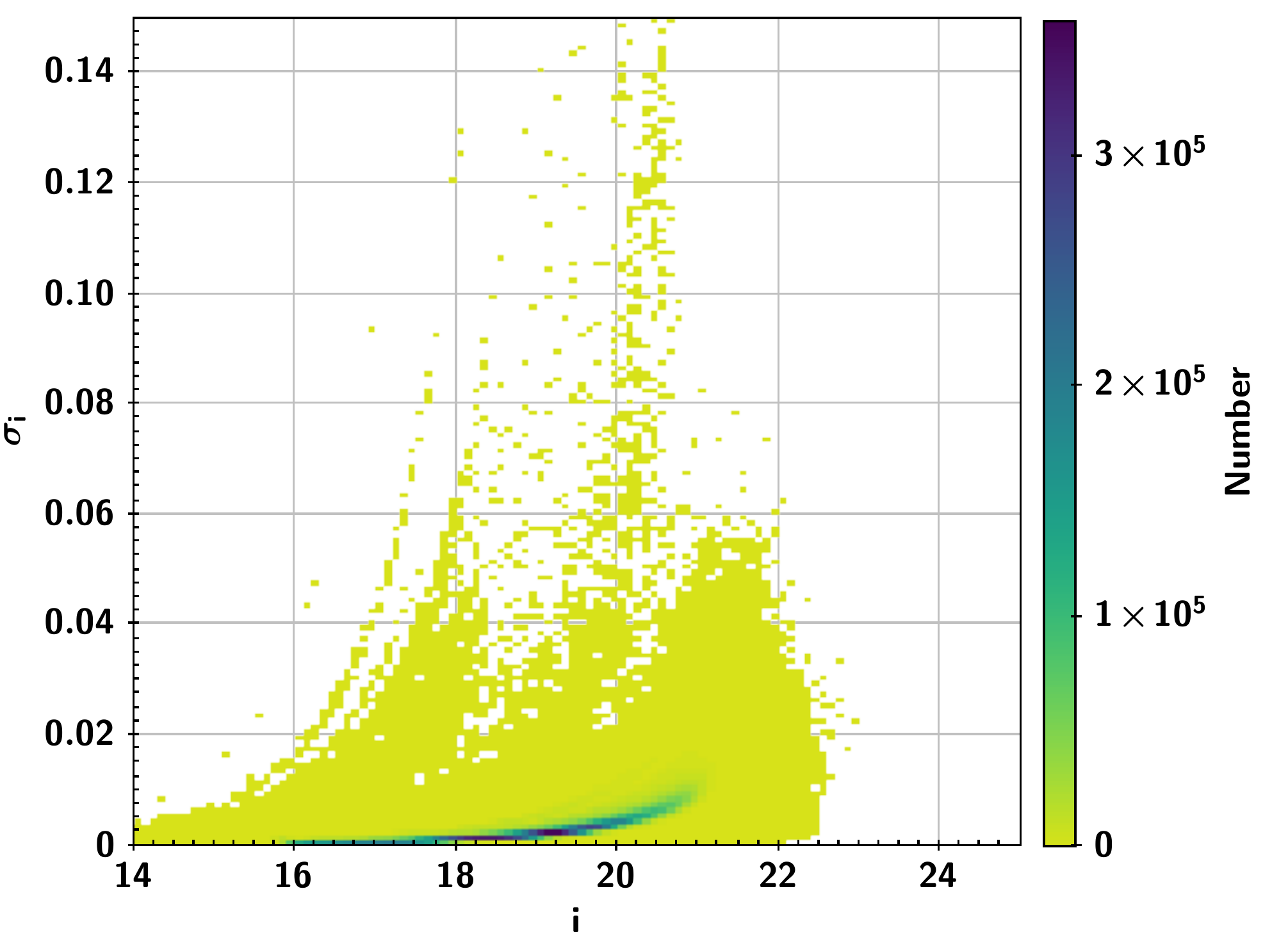}
\caption{$\sigma_i$ vs. $i$.  Data are binned in 0.001~mag ($\sigma_i$) $\times$ 0.1~mag ($i$) bins.  (Here we do not apply any $N_{\rm good}$ cuts.)}
\label{fig:magerr_vs_mag_i}
\end{figure*}

\begin{figure*}[h]
\centering
\includegraphics[scale=0.60]{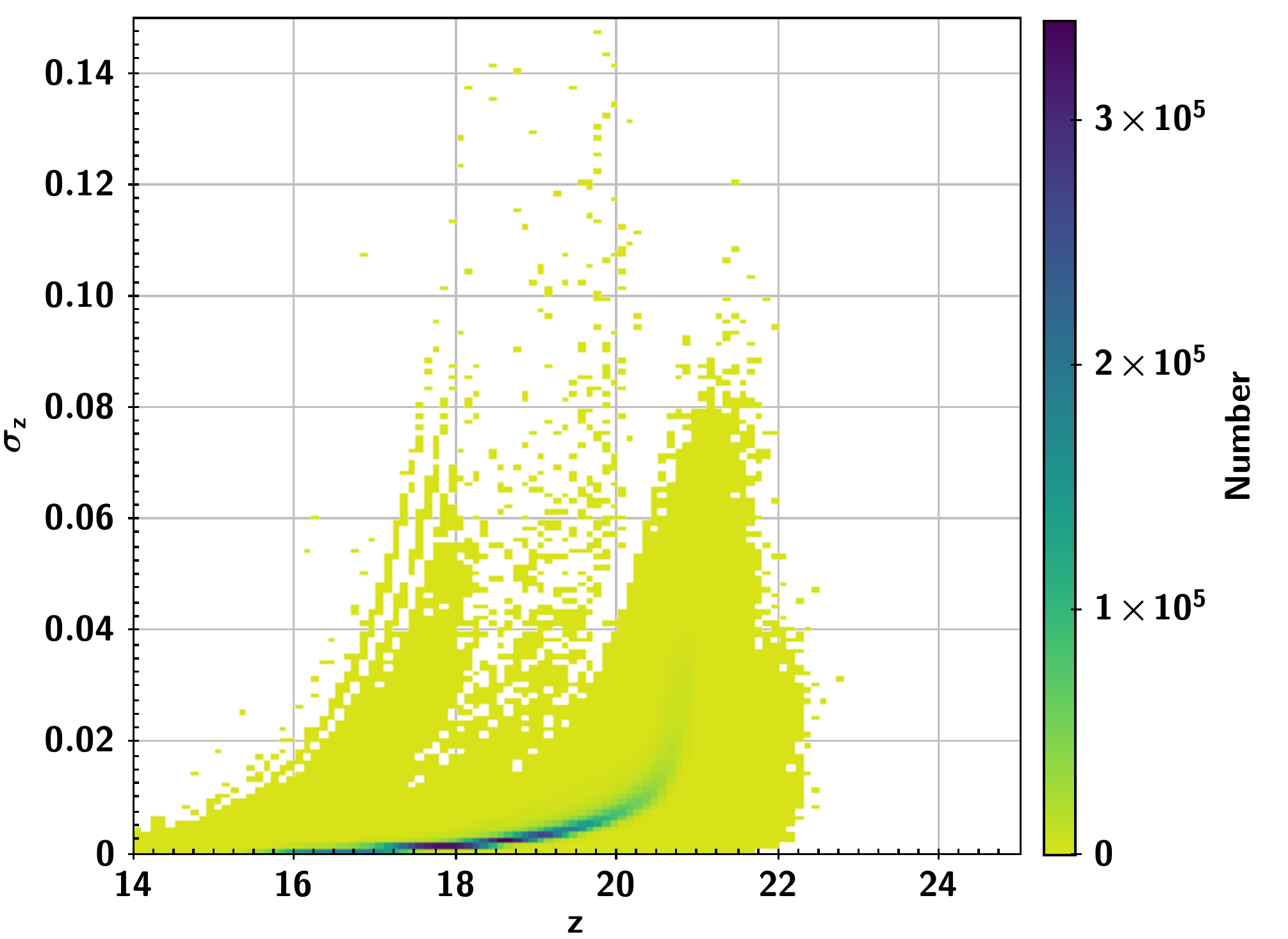}
\caption{$\sigma_z$ vs. $z$.  Data are binned in 0.001~mag ($\sigma_z$) $\times$ 0.1~mag ($z$) bins.  (Here we do not apply any $N_{\rm good}$ cuts.)}
\label{fig:magerr_vs_mag_z}
\end{figure*}

\begin{figure*}[h]
\centering
\includegraphics[scale=0.60]{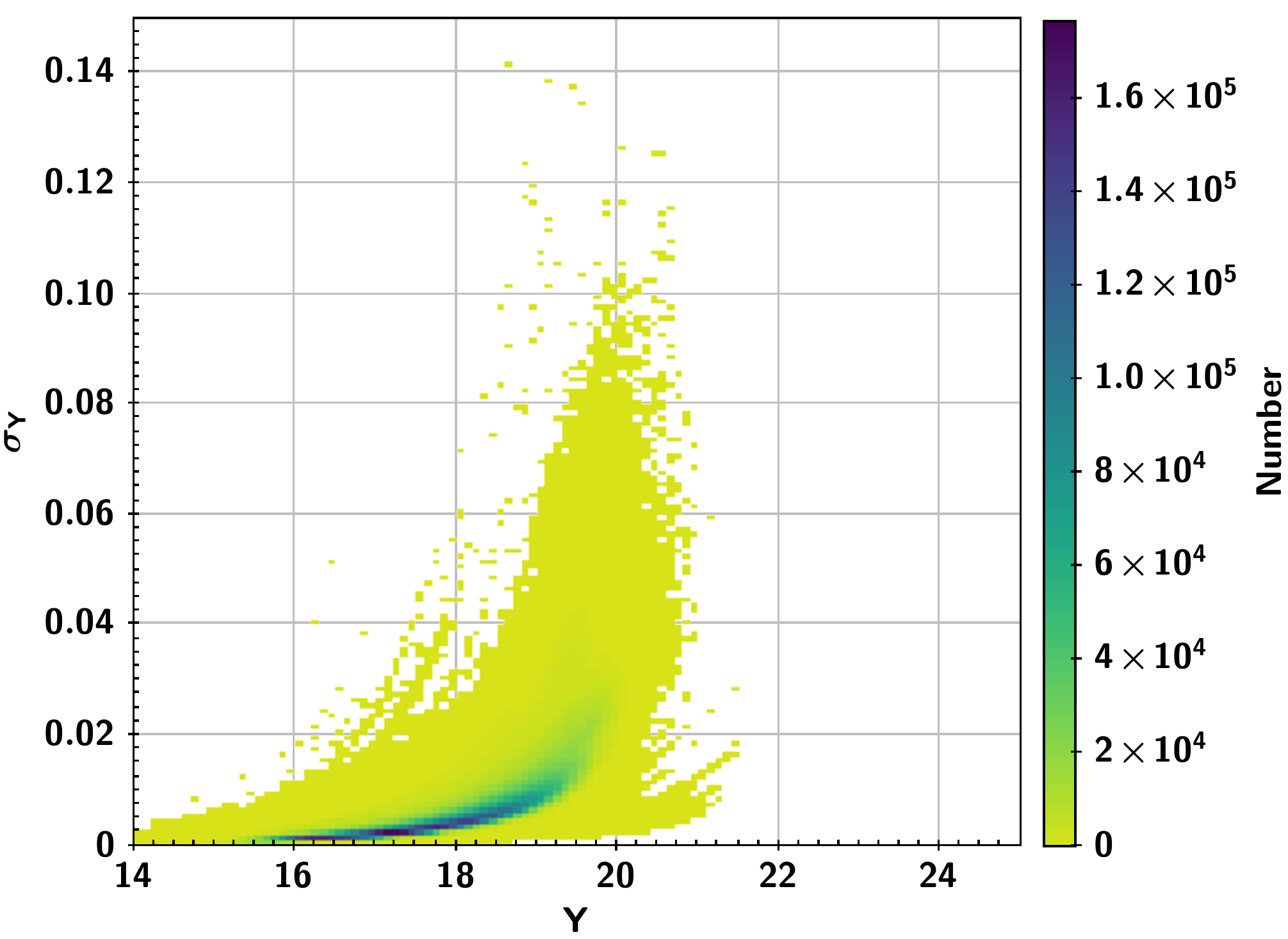}
\caption{$\sigma_Y$ vs. $Y$.  Data are binned in 0.001~mag ($\sigma_Y$) $\times$ 0.1~mag ($y$) bins.  (Here we do not apply any $N_{\rm good}$ cuts.)}
\label{fig:magerr_vs_mag_Y}
\end{figure*}

\begin{figure*}[h]
\centering
\includegraphics[scale=0.70]{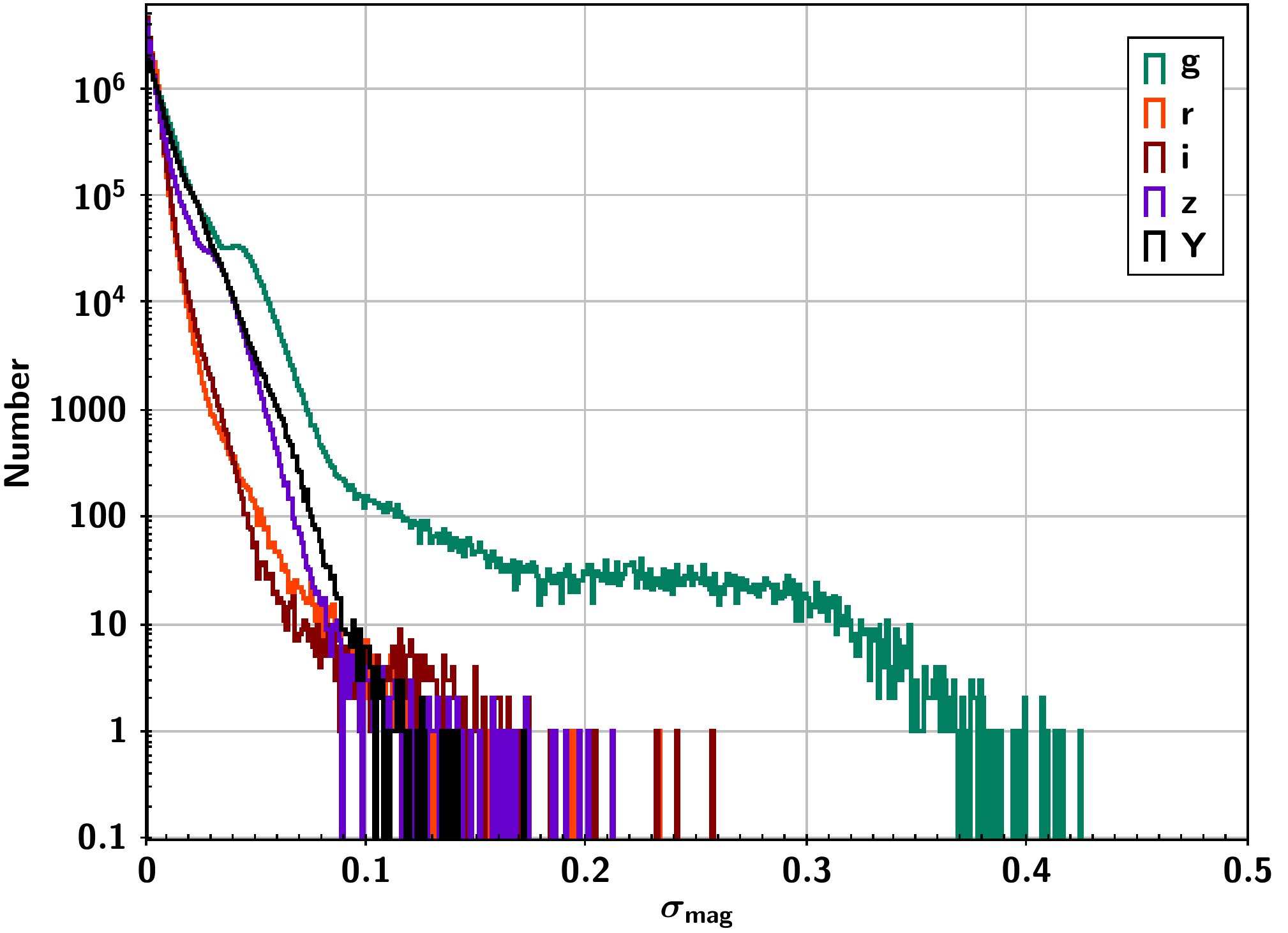}
\caption{Histogram of $\sigma_{\rm mag}$.  Data are binned in 0.001~mag bins.  Note that variable stars and QSOs are not pre-culled from the FGCM calibration star sample, which helps explains the long tail in the distribution of $\sigma_{\rm mag}$'s.  Despite the long tail, over 90\% of the stars have $\sigma_g < 0.02$~mag .  (Here we do not apply any $N_{\rm good}$ cuts.)}
\label{fig:magerr_hist}
\end{figure*}

In addition to the recommendation that the user apply a quality cut on the number of good observations a star has in each filter (preferably $N_{\rm good} > 2$ for a given filter bandpass), we also provide a caveat that, although the vast majority of stars in the catalog have highly precise and accurate calibrated magnitudes, a small fraction have relatively large statistical errors.  This is highlighted for each filter bandpass in Figures~\ref{fig:magerr_vs_mag_g}--\ref{fig:magerr_vs_mag_Y}.  Depending on the use, one may wish to either weight the stars accordingly, or even just cull those stars with high statistical errors.  As seen in the log-linear histograms in Figure~\ref{fig:magerr_hist}, over 90\% of the stars have statistical errors in their photometry of $\sigma_g < 0.02$ (and similar or better in the other filter bands). 

\begin{figure*}[h]
\centering
\includegraphics[scale=0.55]{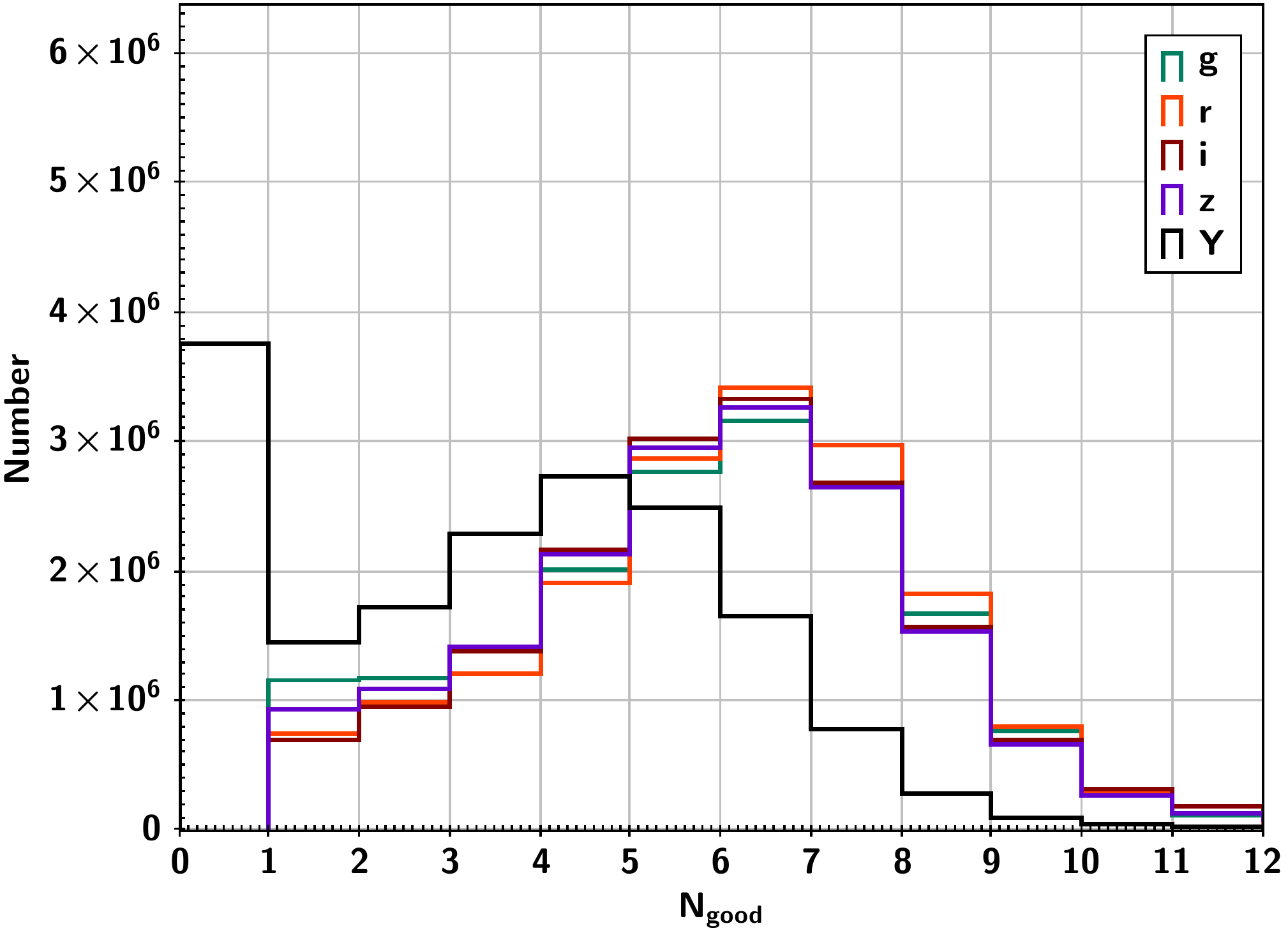}
\caption{Histogram of the number of good measurements that went into the magnitude estimate for each star in a given filter band.  We cut this histogram off at $N_{\rm good} = 12$, which is a reasonable cutoff point for stars outside the 10 DES TF's (a.k.a. the ``supernova'' or ``deep'' fields). Note that a significant number of stars have $N_{\rm good}=0$ for $Y$-band.  Observations made in $Y$-band were not used in the determination of nightly atmospheric parameters, but rather calibrations in $Y$ were ``dead-reckoned'' from observations taken on the same night in other bands (primarily these were $z$-band observations). (Here we do not apply any $N_{\rm good}$ cuts.)}
\label{fig:mag_hist_lin}
\end{figure*}

\begin{figure*}[h]
\centering
\includegraphics[scale=0.55]{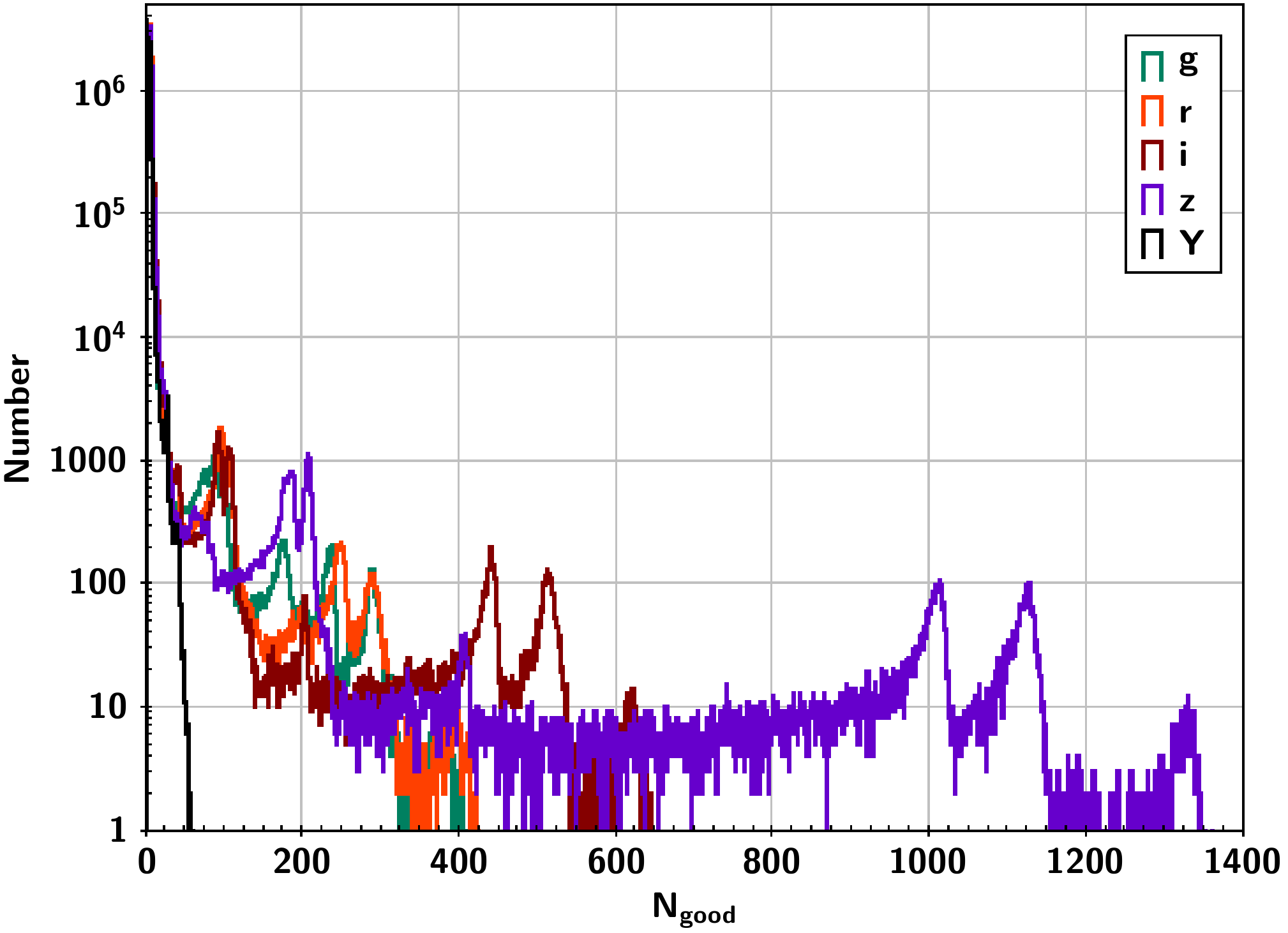}
\caption{Same as previous figure, but the y-axis is now logarithmic.  Note that stars in WF portion of the DES have typically $<$10 good measurements, but that stars in the TF's (the ``deep'' or ``supernova'' fields), which have a much more frequent cadence, have hundreds of observations in the $g$, $r$, $i$, and $z$ bands.  (Here we do not apply any $N_{\rm good}$ cuts.)}
\label{fig:mag_hist_log}
\end{figure*}

So, typically how many good observations were obtained for each star in each filter?  The DES is composed of two parts:  a wide-field (WF) survey covering $\approx5000\,\mathrm{deg}^2$, plus a set of ten 2$.7\,\mathrm{deg}^2$ deep transient fields (TFs; also known as the "supernova fields") that were observed roughly once a week throughout the DES seasons (typically August--February)~\citep{Diehl:2019, 2020arXiv201212824H}.  The goal was to cover the WF portion of the survey with a total of 10 dithered ``tilings'' in $g$,$r$,$i$,$z$ (and, in the end, 8 for $Y$).  Of course, due to the dither pattern, some stars might be observed more than 10 times, and others fewer than 10.  Only photometric observations\footnote{As previously noted, 80\% of the exposures are considered "photometric" in that the variations are modeled by our atmosphere model.} are counted as ``good'' for the $N_{\rm good}$ tabulation. In addition, there are further cuts to individual observations that are deemed outliers.  In Figure~\ref{fig:mag_hist_lin}, we plot the histogram of $N_{\rm good}$ using a linear y-axis and cut off at $N_{\rm good}=12$.  This is appropriate for stars in the WF survey.  In Figure~\ref{fig:mag_hist_log}, we plot the histogram of $N_{\rm good}$ using a logarithmic y-axis and cut off at $N_{\rm good}=12$.  This is appropriate for stars in the TFs.  Note that there are some stars with nearly 1400 good observations in $z$-band.

So far, we have been focused on the uniformity and precision of this catalog.  Another question that remains is how accurately is it tied to a system of physical units (e.g., ergs~s$^{-1}$~cm$^{-2}$~Hz$^{-1}$) traced to the {\em System Internationale} (SI)?  The DES ties itself to the SI by tying itself to the AB magnitude system (\citealt{1983ApJ...266..713O, 1996AJ....111.1748F}); so another way to put this question is,  ``How well is the DES tied to the AB magnitude system?''  To tie this catalog of calibration stars (and hence the DES Year 6 data) to the AB system, FGCM used the synthetic DES $grizY$ AB magnitudes of the star C26202 -- a star that lies in the DES itself and is faint enough not to be saturated in photometric DES science images -- to zeropoint the whole FGCM DES Year 6 calibration star catalog to the AB system.  These synthetic magnitudes were calculated by integrating the official DES passband throughputs (Fig.~\ref{fig:DESDR2_std_bandpasses}) with one of standard spectra for C26202 from the Hubble Space Telescope (HST) CalSpec database \citep{2014PASP..126..711B} -- in this case, the then-current spectrum \var{c26202.stisnic.007}.  

In Table~\ref{tab:c26202} we summarize our results.  For each DES band, we tabulate the synthetic AB magnitude of the \var{C26202.stisnic.007} spectrum (the input values to FGCM), the measured AB magnitudes for C26202 from the final catalog of calibrated stars from FGCM (FGCM mag), the statistical error for the AB magnitude output by FGCM (FGCM $\sigma_{\rm mag}$), the number of good DES observations of C26202 ($N_{\rm good}$), how much the FGCM output magnitude differs from the synthetic DES AB magnitude of \var{C26202.stisnic.007} (AB offset), an estimate of how well the FGCM output magnitude is tied to the synthetic DES AB magnitude of \var{C26202.stisnic.007} ($\sigma_{AB, {\rm stat}}$), and a rough estimate of how well the synthetic DES AB magnitude of \var{C26202.stisnic.007} itself may be tied to the CalSpec AB system ($\sigma_{AB, {\rm sys}}$).  Note that the absolute value of the AB offset is $\le$3~mmag ($\le$0.3\%) for all bands, and that the AB offset is within $2\sigma_{AB, {\rm stat}}$ for all 5 DES passbands, indicating that this catalog is well tied to the synthetic DES AB photometry of \var{C26202}. That said, estimates of how well \var{C26202.stisnic.007} (or the majority of other CalSpec stars) is tied to the AB system is roughly at the 11-12~mmag level (1.1-1.2\% level), as is shown in the final column of Table~\ref{tab:c26202} ($\sigma_{AB, {\rm sys}}$), and this is confirmed by the differences in the synthetic DES AB magnitudes from other versions of the HST CalSpec spectra (see, e.g., \S~4.2.2 and Appendix~C of \citealt{DESDR2}).  Those seeking sub-percent absolute calibrations should note this caveat.

Finally, Table~\ref{tab:calibs_star_cat} provides a listing of the first 25 entries from the DES Y6 FGCM calibration star catalog sorted by RA.  The columns include a running $\var{FGCM\_ID}$ (out of order here, since here we show the catalog after sorting it by ascending RA), the RA and DEC in degrees in J2000.0 coordinates, a \var{FLAG} to note if the star in question was used in the FGCM modeling fit (0) or if the star was randomly excluded from the fit to help with statistical measures (16)\footnote{Roughly 10\% of stars at high Galactic latitudes were reserved, and somewhat fewer at low Galactic latitudes.  Both $\var{FLAG}=0$ and $\var{FLAG}=16$ stars are useful as calibration stars.}, the calibrated DES AB magnitudes of the star in each DES band ($g$, $r$, $i$, $z$, $Y$), the statistical errors in those calibrated magnitudes ($\sigma_{g,r,i,z,Y}$), and the number of good DES observations that went into those calibrated magnitudes ($N_{g,r,i,z,Y}$).  The full machine-readable table can be found at \url{https://des.ncsa.illinois.edu/releases/other}.



\begin{deluxetable}{l c c c r c c c }
\tablewidth{0pt}
\tabletypesize{\tablesize}
\tablehead{
\\
band & \var{stisnic.007} & FGCM mag & FGCM $\sigma_{\rm mag}$ & $N_{\rm good}$  & AB offset  &  $\sigma_{AB, {\rm stat}}$\tablenotemark{\tiny a}  &  $\sigma_{AB, {\rm sys}}$\tablenotemark{\tiny b} \\
     & [mag]                & [mag]                & [mag]  &                & [mag]      &  [mag]                      & [mag] \\ 
}
\tablecaption{Comparison of DES synthetic AB magnitudes and DES observed FGCM PSF magnitudes for the HST CALSPEC Standard C26202 
\label{tab:c26202}}
\startdata
\\
$g$  &  16.695913  &  16.6949 &  0.0001 &  119 & +0.0010  &  0.0018 & 0.011   \\
$r$  &  16.340208  &  16.3432 &  0.0004 &   39 & -0.0030  &  0.0018 & 0.011   \\
$i$  &  16.257366  &  16.2588 &  0.0004 &   38 & -0.0014  &  0.0018 & 0.011   \\
$z$  &  16.245108  &  16.2432 &  0.0001 &  169 & +0.0019  &  0.0018 & 0.012   \\
$Y$  &  16.267345  &  16.2701 &  0.0005 &   49 & -0.0028  &  0.0019 & 0.012   \\
\\
\enddata
\tablenotetext{\tiny a}{The given passband's FGCM $\sigma_{\rm mag}$ added in quadrature with the estimated FGCM sample uniformity vs. Gaia (1.8~mmag).}
\tablenotetext{\tiny b}{A rough estimate of how well DES synthetic magnitudes for C26202 are tied to the true AB system, based on uncertainties in the CalSpec system (see \S~4.2.2 of \citealt{DESDR2}).}
\end{deluxetable}

\movetableright=0.25cm
\begin{rotatetable}
\begin{deluxetable}
{lccrrrrrrrrrrrrrrrr}
\tablewidth{0pt}
\tabletypesize{\scriptsize}
\tablehead{
\\
FGCM\_ID & RA & DEC & FLAG & $g$ & $r$ & $i$ & $z$ & $Y$ & $\sigma_g$ & $\sigma_r$ & $\sigma_i$ & $\sigma_z$ & $\sigma_Y$ & $N_g$ & $N_r$ & $N_i$ & $N_z$ & $N_Y$ \\
              & [deg]    & [deg]     &            &                 &                 &                 &                 &                 &                    &                    &                  &          &           &            &                 &                 &                 &                \\
}
\tablecaption{First 25 entries from the DES Y6 FGCM calibration star catalog (sorted by RA).
\label{tab:calibs_star_cat}}
\startdata
\\
   8200396 &  195.960488 &  -23.418628 &    0 &  20.33161 &  20.04067 &  19.64882 &  19.68032 &  99.00000 &   0.00491 &   0.00598 &   0.00481 &   0.01089 &  99.00000 &     4 &     2 &     3 &     1 &     0 \\ 
   8200420 &  195.960909 &  -23.456651 &    0 &  17.23605 &  16.64793 &  16.43824 &  16.32005 &  16.31429 &   0.00076 &   0.00095 &   0.00060 &   0.00049 &   0.00121 &     5 &     2 &     5 &     9 &     3 \\ 
   8200376 &  195.964547 &  -23.449972 &    0 &  20.41214 &  20.38207 &  20.12734 &  20.11771 &  99.00000 &   0.00339 &   0.00496 &   0.00346 &   0.00662 &  99.00000 &     8 &     4 &    12 &     7 &     0 \\ 
   8200380 &  195.965884 &  -23.393623 &    0 &  21.50839 &  20.63410 &  20.24803 &  20.04050 &  19.94555 &   0.00769 &   0.00602 &   0.00432 &   0.00497 &   0.01712 &     8 &     4 &     9 &    11 &     2 \\ 
   8193725 &  195.966800 &  -23.235245 &    0 &  21.86613 &  21.10612 &  20.82457 &  20.63362 &  99.00000 &   0.01231 &   0.00874 &   0.00976 &   0.01937 &  99.00000 &     5 &     4 &     4 &     2 &     0 \\ 
   8200536 &  195.967935 &  -23.364772 &    0 &  17.98238 &  17.02529 &  16.68442 &  16.49377 &  16.45173 &   0.00080 &   0.00086 &   0.00041 &   0.00040 &   0.00040 &     8 &     3 &    12 &    15 &    29 \\ 
   8193724 &  195.969312 &  -23.203479 &    0 &  21.27325 &  20.00675 &  19.45145 &  19.18036 &  19.12033 &   0.00635 &   0.00376 &   0.00210 &   0.00210 &   0.00270 &     8 &     4 &    11 &    15 &    23 \\ 
   8200054 &  195.970494 &  -23.359857 &    0 &  20.65138 &  20.18068 &  20.01230 &  19.93955 &  99.00000 &   0.00402 &   0.00427 &   0.00336 &   0.00398 &  99.00000 &     8 &     4 &    11 &    15 &     0 \\ 
   8209232 &  195.971048 &  -23.183766 &    0 &  20.58051 &  20.49983 &  20.45573 &  20.13263 &  99.00000 &   0.00404 &   0.00543 &   0.00994 &   0.00639 &  99.00000 &     7 &     4 &     2 &     7 &     0 \\ 
   8192697 &  195.971329 &  -23.243826 &   16 &  20.44195 &  19.99510 &  19.83071 &  19.77511 &  19.79179 &   0.00346 &   0.00373 &   0.00271 &   0.00345 &   0.00886 &     8 &     4 &    12 &    15 &     7 \\ 
   8192690 &  195.972234 &  -23.223584 &    0 &  22.13167 &  20.72738 &  19.97878 &  19.59855 &  19.50391 &   0.03309 &   0.00647 &   0.00327 &   0.00297 &   0.00468 &     1 &     4 &    10 &    15 &    15 \\ 
   8190416 &  195.973693 &  -23.414804 &    0 &  21.86141 &  20.53839 &  19.28752 &  18.74664 &  18.62345 &   0.02374 &   0.00559 &   0.00176 &   0.00149 &   0.00169 &     2 &     4 &    12 &    15 &    27 \\ 
   8190407 &  195.974139 &  -23.359085 &    0 &  21.85823 &  20.34175 &  19.02891 &  18.48030 &  18.33795 &   0.01375 &   0.00481 &   0.00145 &   0.00122 &   0.00134 &     4 &     4 &    12 &    15 &    28 \\ 
   8190412 &  195.974451 &  -23.435365 &    0 &  21.03943 &  20.47173 &  20.29669 &  20.18581 &  99.00000 &   0.00536 &   0.00597 &   0.00399 &   0.00519 &  99.00000 &     8 &     3 &    12 &    13 &     0 \\ 
   8190400 &  195.975022 &  -23.474810 &    0 &  21.00512 &  19.51669 &  18.28084 &  17.73279 &  17.59911 &   0.00546 &   0.00266 &   0.00091 &   0.00073 &   0.00077 &     7 &     4 &    11 &    15 &    30 \\ 
   8190414 &  195.975065 &  -23.460149 &    0 &  19.61247 &  18.42107 &  17.91267 &  17.62720 &  17.56043 &   0.00197 &   0.00135 &   0.00070 &   0.00069 &   0.00077 &     8 &     4 &    12 &    15 &    28 \\ 
   8192799 &  195.977459 &  -23.258257 &    0 &  21.94998 &  20.60057 &  19.81734 &  19.45509 &  19.35236 &   0.01400 &   0.00587 &   0.00292 &   0.00276 &   0.00351 &     4 &     4 &    10 &    13 &    20 \\ 
   8190257 &  195.979145 &  -23.205660 &   16 &  22.17178 &  20.72395 &  19.88035 &  19.48346 &  19.40551 &   0.02628 &   0.00645 &   0.00294 &   0.00269 &   0.00324 &     2 &     4 &    11 &    15 &    26 \\ 
   8200799 &  195.980241 &  -23.453360 &    0 &  19.02387 &  17.63684 &  16.85970 &  16.46340 &  16.36673 &   0.00137 &   0.00092 &   0.00044 &   0.00039 &   0.00038 &     8 &     4 &    12 &    15 &    29 \\ 
   8193739 &  195.981487 &  -23.240127 &    0 &  20.99388 &  20.52345 &  20.35224 &  20.29814 &  99.00000 &   0.00517 &   0.00553 &   0.00497 &   0.00603 &  99.00000 &     8 &     4 &     8 &    11 &     0 \\ 
   8197241 &  195.982043 &  -23.464698 &    0 &  21.76920 &  20.71170 &  20.30091 &  20.03086 &  99.00000 &   0.01197 &   0.00639 &   0.00399 &   0.00443 &  99.00000 &     5 &     4 &    12 &    14 &     0 \\ 
   8193768 &  195.982346 &  -23.272082 &    0 &  20.90443 &  20.47908 &  20.34181 &  20.25568 &  99.00000 &   0.00483 &   0.00534 &   0.00445 &   0.00561 &  99.00000 &     8 &     4 &    10 &    13 &     0 \\ 
   8193740 &  195.982353 &  -23.253331 &    0 &  20.16415 &  19.79998 &  19.66448 &  19.60806 &  19.62355 &   0.00285 &   0.00324 &   0.00256 &   0.00299 &   0.00569 &     8 &     4 &    10 &    15 &    11 \\ 
   8197232 &  195.982849 &  -23.365279 &    0 &  22.33664 &  21.33529 &  20.91051 &  20.69025 &  99.00000 &   0.03956 &   0.01053 &   0.01402 &   0.01269 &  99.00000 &     1 &     4 &     3 &     5 &     0 \\ 
   8197234 &  195.982867 &  -23.419028 &    0 &  19.74854 &  19.32932 &  19.17715 &  19.12428 &  19.13122 &   0.00216 &   0.00235 &   0.00163 &   0.00205 &   0.00255 &     8 &     4 &    12 &    14 &    27 \\ 
\\
\enddata
\tablecomments{ The full machine-readable table can be found at \url{https://des.ncsa.illinois.edu/releases/other} }
\end{deluxetable}
\end{rotatetable}
\newpage


\section{Conclusions}
\label{sec:conclusions}
In this Technical Note, we have presented and described a catalog of roughly 17 million calibrated reference stars, calibrated in the AB magnitude system for the DES $grizY$ bands.  This reference catalog, generated by the Forward Calibration Method (FGCM) pipeline \citepalias{2018AJ....155...41B}, was used for the photometric calibration of the DES 6-Year data set (which itself was the basis for DES DR2; \citealt{DESDR2}).  Due to its large number of stars, its precision and accuracy, and it large coverage of the Southern Sky (approximately $5000\,\mathrm{deg}^2$), we believe this catalog of reference stars will be useful for the photometric calibration of other data -- both DECam data and data using filter sets similar to that of the DES.

The full machine-readable table can be found at \url{https://des.ncsa.illinois.edu/releases/other}.

\acknowledgements

We would like to acknowledge the dedicated work of Dr. Arlo Landolt, and all that he provided to the astronomical community during his lifetime.

Funding for the DES Projects has been provided by the U.S. Department of Energy, the U.S. National Science Foundation, the Ministry of Science and Education of Spain, 
the Science and Technology Facilities Council of the United Kingdom, the Higher Education Funding Council for England, the National Center for Supercomputing 
Applications at the University of Illinois at Urbana-Champaign, the Kavli Institute of Cosmological Physics at the University of Chicago, 
the Center for Cosmology and Astro-Particle Physics at the Ohio State University,
the Mitchell Institute for Fundamental Physics and Astronomy at Texas A\&M University, Financiadora de Estudos e Projetos, 
Funda{\c c}{\~a}o Carlos Chagas Filho de Amparo {\`a} Pesquisa do Estado do Rio de Janeiro, Conselho Nacional de Desenvolvimento Cient{\'i}fico e Tecnol{\'o}gico and 
the Minist{\'e}rio da Ci{\^e}ncia, Tecnologia e Inova{\c c}{\~a}o, the Deutsche Forschungsgemeinschaft and the Collaborating Institutions in the Dark Energy Survey. 

The Collaborating Institutions are Argonne National Laboratory, the University of California at Santa Cruz, the University of Cambridge, Centro de Investigaciones Energ{\'e}ticas, 
Medioambientales y Tecnol{\'o}gicas-Madrid, the University of Chicago, University College London, the DES-Brazil Consortium, the University of Edinburgh, 
the Eidgen{\"o}ssische Technische Hochschule (ETH) Z{\"u}rich, 
Fermi National Accelerator Laboratory, the University of Illinois at Urbana-Champaign, the Institut de Ci{\`e}ncies de l'Espai (IEEC/CSIC), 
the Institut de F{\'i}sica d'Altes Energies, Lawrence Berkeley National Laboratory, the Ludwig-Maximilians Universit{\"a}t M{\"u}nchen and the associated Excellence Cluster Universe, 
the University of Michigan, NFS's NOIRLab, the University of Nottingham, The Ohio State University, the University of Pennsylvania, the University of Portsmouth, 
SLAC National Accelerator Laboratory, Stanford University, the University of Sussex, Texas A\&M University, and the OzDES Membership Consortium.

Based in part on observations at Cerro Tololo Inter-American Observatory at NSF's NOIRLab (NOIRLab Prop. ID 2012B-0001; PI: J. Frieman), which is managed by the Association of Universities for Research in Astronomy (AURA) under a cooperative agreement with the National Science Foundation.

The DES data management system is supported by the National Science Foundation under Grant Numbers AST-1138766 and AST-1536171.
The DES participants from Spanish institutions are partially supported by MICINN under grants ESP2017-89838, PGC2018-094773, PGC2018-102021, SEV-2016-0588, SEV-2016-0597, and MDM-2015-0509, some of which include ERDF funds from the European Union. IFAE is partially funded by the CERCA program of the Generalitat de Catalunya.
Research leading to these results has received funding from the European Research
Council under the European Union's Seventh Framework Program (FP7/2007-2013) including ERC grant agreements 240672, 291329, and 306478.
We  acknowledge support from the Brazilian Instituto Nacional de Ci\^encia
e Tecnologia (INCT) do e-Universo (CNPq grant 465376/2014-2).

This manuscript has been authored by Fermi Research Alliance, LLC under Contract No. DE-AC02-07CH11359 with the U.S. Department of Energy, Office of Science, Office of High Energy Physics.

This work made use of the Illinois Campus Cluster, a computing resource
that is operated by the Illinois Campus Cluster Program (ICCP) in conjunction
with the National Center for Supercomputing Applications (NCSA) and which is
supported by funds from the University of Illinois at Urbana-Champaign.

This research is part of the Blue Waters sustained-petascale computing project,
which is supported by the National Science Foundation (awards OCI-0725070 and
ACI-1238993) and the state of Illinois. Blue Waters is a joint effort of the
University of Illinois at Urbana-Champaign and its National Center for
Supercomputing Applications.

The Science Server described here was developed and is operated by LIneA (Laborat\'orio Interinstitucional de e-Astronomia).

We acknowledge support from the Australian Research Council through project numbers CE110001020, DP160100930, and FL180100168 and the Brazilian Instituto Nacional de Ci\^encia
e Tecnologia (INCT) e-Universe (CNPq grant 465376/2014-2).

This research uses services or data provided by the Astro Data Lab at NSF's National Optical-Infrared Astronomy Research Laboratory. NOIRLab is operated by the Association of Universities for Research in Astronomy (AURA), Inc. under a cooperative agreement with the National Science Foundation.

This work has made use of data from the European Space Agency (ESA) mission
{\it Gaia} (\url{https://www.cosmos.esa.int/gaia}), processed by the {\it Gaia}
Data Processing and Analysis Consortium (DPAC,
\url{https://www.cosmos.esa.int/web/gaia/dpac/consortium}). Funding for the DPAC
has been provided by national institutions, in particular the institutions
participating in the {\it Gaia} Multilateral Agreement.

Funding for the SDSS and SDSS-II has been provided by the Alfred P. Sloan Foundation, the Participating Institutions, the National Science Foundation, the U.S. Department of Energy, the National Aeronautics and Space Administration, the Japanese Monbukagakusho, the Max Planck Society, and the Higher Education Funding Council for England. The SDSS Web Site is \url{http://www.sdss.org/}.
The SDSS is managed by the Astrophysical Research Consortium for the Participating Institutions. The Participating Institutions are the American Museum of Natural History, Astrophysical Institute Potsdam, University of Basel, University of Cambridge, Case Western Reserve University, University of Chicago, Drexel University, Fermilab, the Institute for Advanced Study, the Japan Participation Group, Johns Hopkins University, the Joint Institute for Nuclear Astrophysics, the Kavli Institute for Particle Astrophysics and Cosmology, the Korean Scientist Group, the Chinese Academy of Sciences (LAMOST), Los Alamos National Laboratory, the Max-Planck-Institute for Astronomy (MPIA), the Max-Planck-Institute for Astrophysics (MPA), New Mexico State University, Ohio State University, University of Pittsburgh, University of Portsmouth, Princeton University, the United States Naval Observatory, and the University of Washington.

Funding for SDSS-III has been provided by the Alfred P. Sloan Foundation, the Participating Institutions, the National Science Foundation, and the U.S. Department of Energy Office of Science. The SDSS-III web site is \url{http://www.sdss3.org/}.
SDSS-III is managed by the Astrophysical Research Consortium for the Participating Institutions of the SDSS-III Collaboration including the University of Arizona, the Brazilian Participation Group, Brookhaven National Laboratory, Carnegie Mellon University, University of Florida, the French Participation Group, the German Participation Group, Harvard University, the Instituto de Astrofisica de Canarias, the Michigan State/Notre Dame/JINA Participation Group, Johns Hopkins University, Lawrence Berkeley National Laboratory, Max Planck Institute for Astrophysics, Max Planck Institute for Extraterrestrial Physics, New Mexico State University, New York University, Ohio State University, Pennsylvania State University, University of Portsmouth, Princeton University, the Spanish Participation Group, University of Tokyo, University of Utah, Vanderbilt University, University of Virginia, University of Washington, and Yale University.

Funding for the Sloan Digital Sky 
Survey IV has been provided by the 
Alfred P. Sloan Foundation, the U.S. 
Department of Energy Office of 
Science, and the Participating 
Institutions. 
SDSS-IV acknowledges support and 
resources from the Center for High 
Performance Computing  at the 
University of Utah. The SDSS 
website is www.sdss4.org.
SDSS-IV is managed by the 
Astrophysical Research Consortium 
for the Participating Institutions 
of the SDSS Collaboration including 
the Brazilian Participation Group, 
the Carnegie Institution for Science, 
Carnegie Mellon University, Center for 
Astrophysics | Harvard \& 
Smithsonian, the Chilean Participation 
Group, the French Participation Group, 
Instituto de Astrof\'isica de 
Canarias, The Johns Hopkins 
University, Kavli Institute for the 
Physics and Mathematics of the 
Universe (IPMU) / University of 
Tokyo, the Korean Participation Group, 
Lawrence Berkeley National Laboratory, 
Leibniz Institut f\"ur Astrophysik 
Potsdam (AIP),  Max-Planck-Institut 
f\"ur Astronomie (MPIA Heidelberg), 
Max-Planck-Institut f\"ur 
Astrophysik (MPA Garching), 
Max-Planck-Institut f\"ur 
Extraterrestrische Physik (MPE), 
National Astronomical Observatories of 
China, New Mexico State University, 
New York University, University of 
Notre Dame, Observat\'ario 
Nacional / MCTI, The Ohio State 
University, Pennsylvania State 
University, Shanghai 
Astronomical Observatory, United 
Kingdom Participation Group, 
Universidad Nacional Aut\'onoma 
de M\'exico, University of Arizona, 
University of Colorado Boulder, 
University of Oxford, University of 
Portsmouth, University of Utah, 
University of Virginia, University 
of Washington, University of 
Wisconsin, Vanderbilt University, 
and Yale University.

This research makes use of the SciServer science platform (www.sciserver.org). 
SciServer is a collaborative research environment for large-scale data-driven science. It is being developed at, and administered by, the Institute for Data Intensive Engineering and Science at Johns Hopkins University. SciServer is funded by the National Science Foundation through the Data Infrastructure Building Blocks (DIBBs) program and others, as well as by the Alfred P. Sloan Foundation and the Gordon and Betty Moore Foundation.

DLT would like to thank Bob Abel, Christina Adair, Allan Jackson, and others at Rubin Data Preview Assemblies and Rubin Stack Club meetings for productive conversations on spectroscopic classification of celestial objects using SDSS spectra.

\vspace{5mm}
\facility{Blanco (DECam), Astro Data Lab, SciServer} 

\software{
\healpix \citep{HealpixSoft},\footnote{\url{http://healpix.sourceforge.net}}
\code{healpy} \citep{Zonca2019},\footnote{\url{https://github.com/healpy/healpy}}
\code{healsparse},\footnote{\url{https://healsparse.readthedocs.io/en/latest/}}
\code{matplotlib} \citep{2007CSE.....9...90H}, 
\code{numpy} \citep{numpy:2011}, 
\code{scipy} \citep{scipy:2001}, 
\code{astropy} \citep{Astropy:2013}, 
\code{fitsio},\footnote{\url{https://github.com/esheldon/fitsio}}
\code{easyaccess}\citep{easyaccess}, 
\code{skymap},\footnote{\url{https://github.com/kadrlica/skymap}}
\code{TOPCAT} \citep{2005ASPC..347...29T},
\code{STILTS} \citep{2005ASPC..347...29T}
}

\bibliographystyle{yahapj_twoauthor_arxiv_amp}
\bibliography{biblio}

\end{document}